\documentclass[]{jfm}

\usepackage{natbib}
\usepackage{hyperref}
\hypersetup{colorlinks=true,urlcolor=blue,citecolor=black}
\usepackage{pgfplots}

\usepackage{graphicx}
\usepackage{amsmath}
\usepackage{xcolor}
\usepackage{dashrule}
\usepackage{tikz}

\newcommand{\aver}[1]{ \! \left\langle {#1} \right \rangle \!}
\newcommand{\vect}[1]{\boldsymbol{#1}}

\title[Cascades of Reynolds stresses in Couette flow]{Ascending-descending and direct-inverse cascades of Reynolds stresses \\ in turbulent Couette flow}

\author[A. Chiarini, M. Mauriello, D. Gatti and M. Quadrio]
{
\ns A.\ns C\ls H\ls I\ls A\ls R\ls I\ls N\ls I\ls$^1$,
\ns M.\ns M\ls A\ls U\ls R\ls I\ls E\ls L\ls L\ls O\ls$^1$\thanks{Current address: Aix-Marseille Univ., CNRS, IUSTI UMR 7343, 13013 Marseille, France},
\ns D.\ns G\ls A\ls T\ls T\ls I\ls$^2$ 
\and M.\ns Q\ls U\ls A\ls D\ls R\ls  I\ls O\ls$^1$
}

\affiliation{
  $^1$ Dipartimento di Scienze e Tecnologie Aerospaziali del Politecnico di Milano, 
  via La Masa 34, 20156 Milano, Italy\\
  $^2$ Institute of Fluid Mechanics, Karlsruhe Institute of Technology, Kaiserstr. 10, 76131 Karlsruhe, Germany
  \\[\affilskip]
}

\begin{document}
\maketitle

\begin{abstract}

The interaction between small- and large-scale structures, and the coexisting bottom-up and top-down processes are studied in a turbulent plane Couette flow, where space-filling longitudinal rolls appear at relatively low values of the Reynolds number $Re$. A DNS database at $Re_\tau=101$ is built to replicate the highest $Re$ considered in a recent experimental work by Kawata and Alfredsson ({\em Phys. Rev. Lett.}, vol.120, 2018, 244501). Our study is based on the exact budget equations for the second-order structure function tensor $\aver{\delta u_i \delta u_j}$, i.e. the Anisotropic Generalized Kolmogorov Equations (AGKE). The AGKE study production, redistribution, transport and dissipation of every Reynolds stress tensor component, considering simultaneously the physical space and the space of scales, and properly define the concept of scale in the inhomogeneous wall-normal direction.

We show how the large-scale energy-containing motions are involved in the production and redistribution of the turbulent fluctuations. Both bottom-up and top-down interactions occur, and the same is true for direct and inverse cascading. The wall-parallel components $\aver{\delta u \delta u}$ and $\aver{\delta w \delta w}$ show that both small and large near-wall scales feed the large scales away from the wall. The wall-normal component $\aver{\delta v \delta v}$ is different, and shows a dominant top-down dynamics, being produced via pressure-strain redistribution away from the wall and transferred towards near-wall larger scales via an inverse cascade. The off-diagonal component shows a top-down interaction, with both direct and inverse cascade, albeit the latter takes place within a limited range of scales.

\end{abstract}

\section{Introduction}
Provided the value of the Reynolds number $Re$ is not too low, a typical wall-bounded turbulent flow features both large-scale structures located away from but extending down to the wall, and smaller-scale structures which dominate the near-wall region but populate the whole flow. In recent years, the top-down influence of large-scale structures has been thoroughly studied, and considerable efforts have been devoted to their statistical characterisation and to the understanding of the anomalous viscous scaling for the near-wall statistics attributed to the outer structures \citep[see e.g.][]{smits-mckeon-marusic-2011}. On the other hand, how the smaller-scale structures residing near the wall affect the large-scale structures away from the wall is less clear.

Early studies addressing such a bottom-up interaction attributed the existence of large-scale  motions to agglomeration of smaller-scale events \citep{guala-hommema-adrian-2006}, while \cite{toh-itano-2005} introduced the idea of a co-supporting cycle where the large structures are directly forced by the near-wall ones. There are also suggestions \citep{flores-jimenez-delalamo-2007, hwang-cossu-2010} that large-scale motions in the outer layer can self-sustain and are nearly independent from the buffer layer. Such multi-scale interactions have come into focus only recently, one of the reason being that they become visible once the Reynolds number of the flow is large, owing to the required separation of scales. Moreover, they have been mostly studied in the turbulent channel flow, because of its geometrical simplicity. The tool of choice for such studies is often the spectral analysis of the transport equation for the components of the Reynolds stress tensor \citep{mizuno-2016, lee-moser-2019}. A notable result is that in wall flows the (statistically dominant) energy transfer towards small scales is accompanied by an inverse transfer from small to large scales. However, spectral analysis can only observe energy fluxes in the space of scales, and the concept of scale is limited to homogeneous directions. An alternative tool, the Generalized Kolmogorov Equation (GKE), seems well suited for this analysis. The GKE was derived by \cite{hill-2001} as an exact budget equation for the second-order structure function: a separation in the inhomogeneous wall-normal direction can be defined, and a description of energy production, dissipation and energy fluxes in the compound space of scales and positions becomes possible. 

The GKE has been used, in more or less simplified forms, to describe the effects of inhomogeneity on small-scale turbulence, most often for the plane channel flow, although for example \cite{mollicone-etal-2018} used the GKE to identify the coherent vortical structures in a separated turbulent flow behind a bump, \cite{togni-cimarelli-deangelis-2015} to describe Rayleigh--B\'enard convection, \cite{portela-papadakis-vassilicos-2017} to address the turbulent cascades in the near wake of a square cylinder, and \cite{cimarelli-etal-2021} to investigate the entrainment and mixing processes in a turbulent planar temporal jet. \cite{cimarelli-deangelis-casciola-2013} applied the GKE to a turbulent channel flow and observed that energy fluxes originating in the near-wall region transfer energy backwards towards longer and wider turbulent structures as the distance from the wall increases, and then interact with smaller structures before eventually dissipating. In a further study, \cite{cimarelli-etal-2016} identified an additional energy source, which appears in the overlap layer as $Re$ increases and leads to a complex spatial redistribution of energy which links small dissipative and large energy-containing scales through a mixed cascade. An inverse energy transfer from small- to large-scale motions has been also observed near the wall by \cite{cho-hwang-choi-2018} by means of the spectral TKE balance equation. They provided evidence that the large-scale motions near the wall scale in inner units, and \cite{cheng-etal-2020} revealed that the footprints of the large-scale motions manifest as large-scale regions of negative wall friction.

Less attention has been devoted to the plane turbulent Couette flow, where to our knowledge a GKE analysis has never been attempted. However, Couette is as geometrically simple as Poiseuille, and represents an interesting alternative for probing the inner/outer interactions at DNS-accessible Reynolds numbers. In fact, this is one of the most fundamental configurations of wall-bounded turbulence, where two indefinite parallel plates move at constant relative velocity and produce a purely shear-driven turbulence. 

A key difference with the pressure-driven Poiseuille flow is that the mean velocity gradient for Couette does not vanish at the centerline, leading to a non-zero production of turbulent kinetic energy in the core region. Moreover, already at relatively low $Re$, Couette contains large-scale motions in the form of elongated streamwise vortices filling the space between the two walls: for example \cite{tsukahara-kawamura-shingai-2006} found evidence of these vortices already at $Re_\tau=52$, and \cite{lee-moser-2018} described them at $Re_\tau=93$. These structures, traced down by \cite{illingworth-2020} to the linearized Squire and Orr--Sommerfeld operators, can be identified in both flows, but Couette creates stronger structures out of the mean shear \citep{andreolli-quadrio-gatti-2021}. The mutual interaction between the large streamwise-elongated rolls and the smaller eddies in the turbulent flow is not entirely understood. \cite{lee-moser-2019} discussing high-$Re$ Poiseuille flow observed how energy is transferred more or less isotropically from the largest structures down to the dissipation scales via non-linear interactions. However, \cite{kawata-tsukahara-2021} artificially constrained the wall-parallel dimensions of the computational domain for a Couette flow to indirectly characterise direct and inverse cascades, and used data computed from spanwise-minimal domains to suggest that the interscale energy transfers observed by \cite{lee-moser-2019} with one-dimensional spanwise spectral analysis might pertain to the inner and outer dynamics, without actually representing their interactions. \cite{kawata-alfredsson-2018} experimentally studied Couette flow via stereoscopic particle-image velocimetry, and were first to consider the Reynolds shear stress. They observed an inverse cascade from the small scales near the wall to the large scales away from the wall, but only for the shear stress. The inverse cascade mainly affects the near-wall region, and exerts a bottom-up influence from the near-wall region to the channel core. In their experimental study, \cite{kawata-alfredsson-2018} had access to the three velocity components in nine wall-parallel planes (outside the near-wall region), and had to resort to the continuity equation to indirectly assess the wall-normal derivative of the wall-normal velocity component, whereas the remaining wall-normal derivatives were not considered in the transfer. Moreover, in their analysis the streamwise separation was neglected. 

The present work comprehensively addresses the interscale exchanges across the wall-normal direction in a turbulent plane Couette flow, and builds upon a recent extension of the GKE \citep{gatti-etal-2020} that enables the observation of each component of the Reynolds stress tensor in the compound space of scales and wall distances. This extended tool, the Anisotropic Generalized Kolmogorov Equations or AGKE, will be used here to study a Couette flow computed with DNS at $Re_\tau \approx 100$, the highest $Re$ considered by \cite{kawata-alfredsson-2018}. Our aim is to address the mechanisms of direct/inverse cascading, and to identify bottom-up and top-down processes. The paper is organized as follows. First, in \S\ref{sec:methods} the budget equations for the structure function tensor are recalled, and the simulation which produced the DNS database used later for statistical analysis is briefly described. Then, in \S\ref{sec:results} the main results concerning the components of the Reynolds stresses tensor are presented, and in \S\ref{sec:conclusions} a concluding discussion is given. 

\section{Methods}
\label{sec:methods}

\begin{figure}
\centering
\begin{tikzpicture}
\draw [thick, ->, -latex] (10,0) -- (16,2) node [ below ] {$\vect{x'}$};
\draw (10,0) node [below] {$\vect{x}$};
\draw[fill] (13,1) circle [radius=0.05] node [below] {$\vect{X}$};
\draw (14,1.3) node [below] {$\vect{r}$};
\draw [thick, ->, -latex] (16,2) -- (14,4) node [ above ] {$\vect{u' = u(x')}$};
\draw [dashed, thick, ->, -latex] (10,0) -- (8,2) node [below left] {$\vect{u' = u(x')}$};
\draw [thick, ->, -latex] (10,0) -- (11,3) node [ above ] {$\vect{u = u(x)}$};
\draw [red, thick, ->, -latex] (11,3) -- (8,2) node [ above, midway ] {\textcolor{black}{$\vect{\delta u}$}};
\end{tikzpicture}
\caption{Sketch of the quantities involved in the definition of the second-order structure function tensor $\aver{\delta u_i \delta u_j}$. Velocities $\vect{u}$ and $\vect{u'}$ are evaluated at the points $\vect{x}$ and $\vect{x'}$ and used to define the increment $\vect{\delta u}$. The midpoint is $\vect{X} = (\vect{x'}+\vect{x})/2$, while the separation vector is $\vect{r} = \vect{x'} - \vect{x}$.}
\label{fig:sketch}
\end{figure}
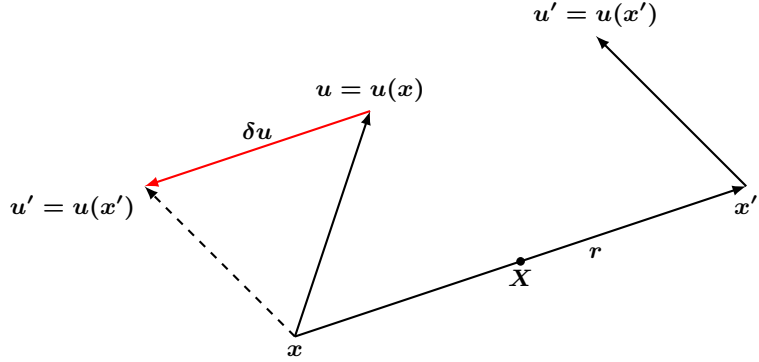

This work relies upon the AGKE budget equations as introduced by \cite{gatti-etal-2020} to tackle the inhomogeneity and the anisotropy of a generic turbulent flow. These equations are briefly recalled here for completeness. 

The AGKE are exact budget equations for the second-order structure function tensor $\aver{\delta u_i \delta u_j}$; they derive from manipulation of the Navier--Stokes equations, and provide a dynamical description of turbulence both in the space of scales and at physical locations. The tensor $\aver{\delta u_i \delta u_j}$ features the components of the increment of the fluctuating velocity vector $\vect{\delta u} = \vect{u}(\vect{x'}) - \vect{u}(\vect{x})$ between two points $\vect{x}$ and $\vect{x'}$; the midpoint and the separation vector are defined as $\vect{X} = (\vect{x'}+\vect{x})/2$ and $\vect{r = x' - x}$, respectively. In its most general form, the tensor $\aver{\delta u_i \delta u_j}$ is a function of seven independent variables, i.e. the six coordinates of the vectors $\vect{X}, \vect{r}$, and the time $t$. By definition, it combines the covariance $V_{ij}$ of the velocity fluctuations and the spatial correlation tensor $R_{ij}$: 
\begin{equation}
\aver{\delta u_i \delta u_j}(\vect{X}, \vect{r},t) = V_{ij} (\vect{X}, \vect{r},t) - R_{ij} (\vect{X}, \vect{r},t) - R_{ij} (\vect{X}, -\vect{r},t)
\end{equation}
where
\begin{equation}
V_{ij} (\vect{X}, \vect{r}, t)  = \aver{u_i u_j} ( \vect{X} + \frac{\vect{r}}{2},t) + \aver{u_i u_j} ( \vect{X} - \frac{\vect{r}}{2},t)
\end{equation} 
is the sum of the covariances $\aver{u_i u_j}$ evaluated at the points $\vect{X} \pm \vect{r}/2$ at time $t$, and
\begin{equation}
R_{ij} (\vect{X}, \vect{r}, t) = \aver{u_i \bigg( \vect{X} + \frac{\vect{r}}{2},t \bigg) u_j \bigg( \vect{X} - \frac{\vect{r}}{2},t \bigg) } 
\end{equation}
is the two-point spatial correlation tensor. For sufficiently large values of $|\vect{r}|$, the correlation vanishes, and $\aver{\delta u_i \delta u_j}$ reduces to $V_{ij}$; thus at large separations the AGKE equations yield the sum of budget equations for the single-point Reynolds stresses at $\vect{X} \pm \vect{r}/2$. 

In the present work, which considers the Couette flow, the AGKE are applied to a statistically stationary flow where only the wall-normal direction $y$ or $x_2$ is statistically inhomogeneous; hence $X$, $Z$ and $t$ drop from the list of independent variables, which reduce to four, i.e. ($r_x,r_y,r_z,Y$). Introducing the four-components vector of the scale and space fluxes $(\phi_{k,ij}, \psi_{ij})$ with $k=1,2,3$, and the source term $\xi_{ij}$, the AGKE are compactly written as: 
\begin{equation}
\frac{\partial \phi_{k,ij}}{\partial r_{k}} + \frac{\partial \psi_{ij}}{\partial Y} = \xi_{ij}
\label{eq:AGKE}
\end{equation}
where repeated indices imply summation. The components of the vector of fluxes are defined as:
\begin{equation}
\phi_{k,ij} = \underbrace{\aver{\delta U \delta u_i \delta u_j}\delta_{k1}}_{\mbox{\tiny mean transport}} \underbrace{+\aver{\delta u_k \delta u_i \delta u_j}}_{\mbox{\tiny turbulent transport}} \underbrace{-2 \nu \frac{\partial}{\partial r_k}\aver{\delta u_i \delta u_j}}_{\mbox{\tiny viscous diffusion}} \qquad  k = 1,2,3 
\label{eq:Phi}
\end{equation}
\begin{equation}
\psi_{ij} = \underbrace{\aver{v^*\delta u_i \delta u_j}}_{\mbox{\tiny turbulent transport}}\underbrace{+\frac{1}{\rho}\aver{\delta p \delta u_j}\delta_{i2}+\frac{1}{\rho}\aver{\delta p \delta u_i}\delta_{j2}}_{\mbox{\tiny pressure transport}}\underbrace{-\frac{\nu}{2}\frac{\partial}{\partial Y}\aver{\delta u_i \delta u_j}}_{\mbox{\tiny viscous diffusion}} \label{eq:Psi}
\end{equation}
and the source term as:
\begin{equation}
\begin{split}
\xi_{ij} = & \underbrace{-\aver{v^* \delta u_j}\delta\bigg( \frac{dU}{dy} \bigg) \delta_{i1} - \aver{v^* \delta u_i}\delta\bigg( \frac{dU}{dy} \bigg) \delta_{j1}}_{\mbox{\tiny production }  (P_{ij}) } + \\
&\underbrace{-\aver{\delta v\delta u_j} \bigg( \frac{dU}{dy} \bigg )^* \delta_{i1} - \aver{\delta v\delta u_i} \bigg( \frac{dU}{dy} \bigg )^* \delta_{j1}}_{\mbox{\tiny production } (P_{ij})} + \\
&\underbrace{+ \frac{1}{\rho}\aver{\delta p \frac{\partial \delta u_i}{\partial X_j}} + \frac{1}{\rho}\aver{\delta p \frac{\partial \delta u_j}{\partial X_i}} }_{\mbox{\tiny pressure strain }(\Pi_{ij})} \underbrace{-4\epsilon_{ij} ^*.}_{\mbox{\tiny ps.dissipation } (D_{ij})} 
\label{eq:xi}
\end{split}
\end{equation}

In the expression above, $\delta_{ij}$ is the Kronecker delta, and the asterisk superscript $f^{*}$ denotes the generic quantity $f$ averaged between the two positions $\vect{X} \pm \vect{r}/2$. The components of the vector of fluxes describe the flux of $\aver{\delta u_i \delta u_j}$ in the space of scales and in the physical space by means of $\phi_{k,ij}$ and $\psi_{ij}$, respectively. In each term, the mean and turbulent transport, the pressure transport and the viscous diffusion are recognized in analogy with the single-point budget equations for the Reynolds stresses $\aver{u_i u_j}$ \citep{pope-2000}. The source term $\xi_{ij}$ describes the net production of $\aver{\delta u_i \delta u_j}$; in addition to production and dissipation, it also features a pressure-strain term, which redistributes turbulent energy among the different components of turbulent stresses.

It is worth pointing out that the AGKE terms possess analytical symmetries or anti-symmetries with respect to an inversion of the separation vector $\vect{r}$; moreover, they additionally enjoy a statistical symmetry (or anti-symmetry) when both the wall-normal and streamwise coordinates are inverted. These symmetries, which are exploited in the numerical code to minimise the computational effort, have been reported by \cite{gatti-etal-2020} for the Poiseuille flow in their Appendix B, and are listed here in Appendix \ref{sec:sym} in the form valid for the Couette flow.

\subsection{The DNS database and the AGKE code}

The statistical analysis discussed in the following stems from the post-processing of a DNS database produced for a turbulent plane Couette flow at moderate Reynolds number. The two channel walls move with speed $\pm U_w$ along the streamwise $x$ direction and are separated in the wall-normal direction $y$ by a gap $L_y=2h$. The Reynolds number is defined as $Re = U_w h / \nu$, where $\nu$ is the kinematic viscosity of the fluid, and is set at $Re= 1666$ corresponding to a value for the Reynolds number based on the friction velocity $u_\tau = \sqrt{\tau_w/\rho}$ of $Re_\tau =101.6$. Throughout this work, all the quantities denoted with the superscript + are given in viscous units, i.e. normalized with $u_\tau$ and $\nu$.

The simulation was carried out with the mixed-discretization DNS code introduced by \cite{luchini-quadrio-2006}, in which the incompressible Navier--Stokes equations are solved in the divergence-free space spawned by the wall-normal velocity $v$ and wall-normal vorticity $\eta$ by means of a pseudo-spectral method. A Fourier discretization is adopted in the homogeneous directions, and fourth-order compact explicit finite-differences schemes are used for the wall-normal derivatives. Temporal integration is partially implicit, with a third-order Runge--Kutta scheme for the explicit convective part and a second-order Crank--Nicolson scheme for the viscous terms treated implicitly. 

The size of the computational domain is $12 \pi h \times 2h \times 4 \pi h$ ($L_x$, $L_y$ and $L_z$ respectively) in the streamwise, wall-normal and spanwise directions. The wall-parallel directions are discretized with $N_x = N_z = 384$ Fourier modes, including those required to exactly remove the aliasing error according to the 3/2 rule. In the wall-normal direction a hyperbolic tangent distribution for the $N_y = 128$ points is used to obtain a more refined grid near the wall. The simulation is led to statistical equilibrium and then advanced in time for further $1500 h / U_w$, during which one hundred equally spaced flow fields are stored for further analysis.

\begin{figure}
\centering
\includegraphics[width=0.8\textwidth]{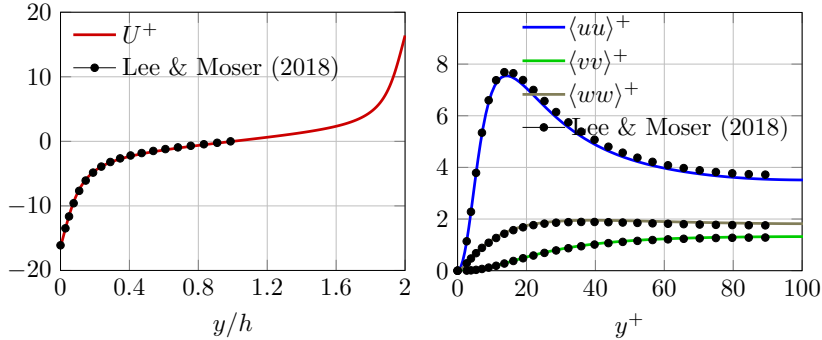}
\caption{Comparison between the present results (color lines) and those by \cite{lee-moser-2018} (black symbols), obtained at a slightly different $Re$. Left: mean velocity profile; right: diagonal Reynolds stresses.}
\label{fig:profiles}
\end{figure}

\begin{figure}
\centering
\includegraphics[width=0.8\textwidth]{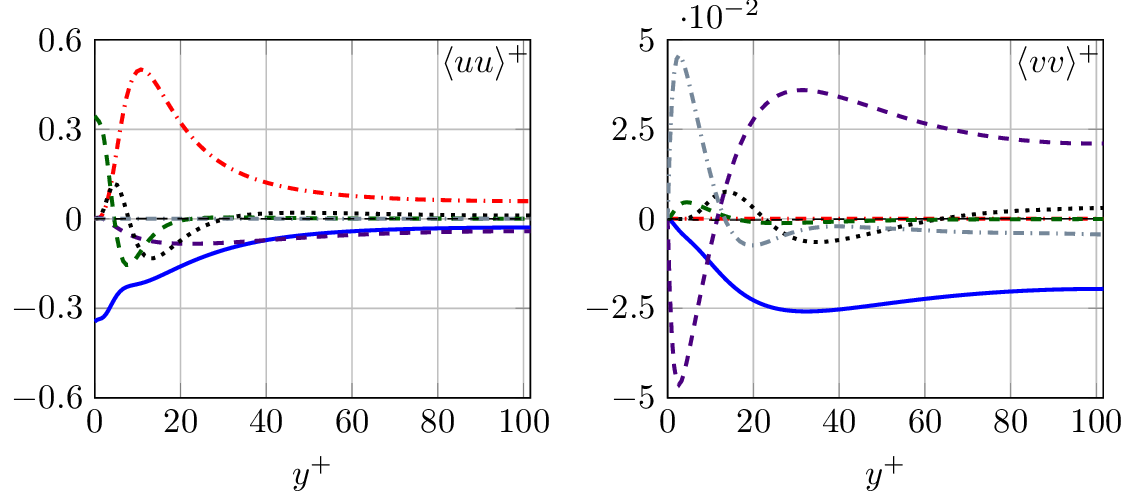}
\includegraphics[width=0.8\textwidth]{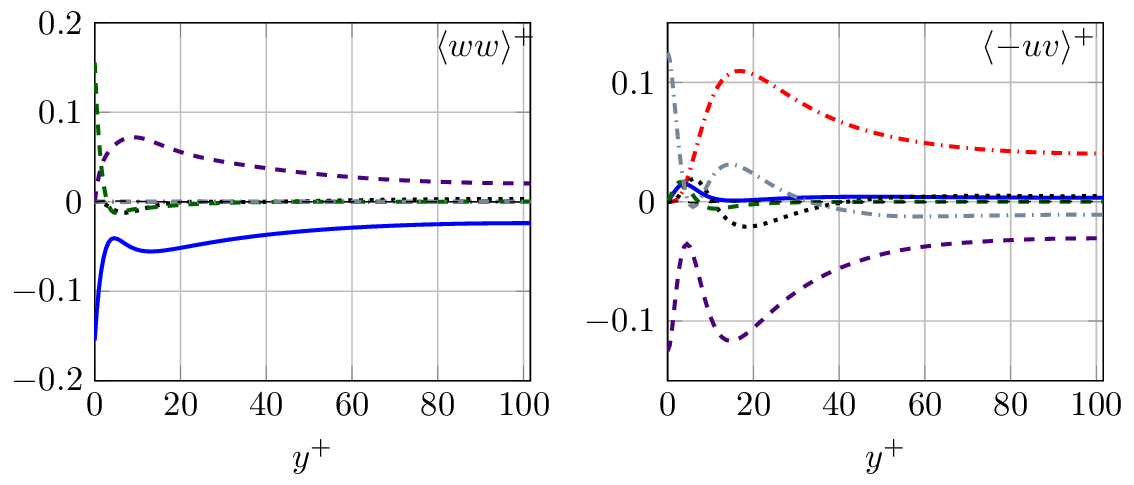}
\includegraphics[width=0.4\textwidth]{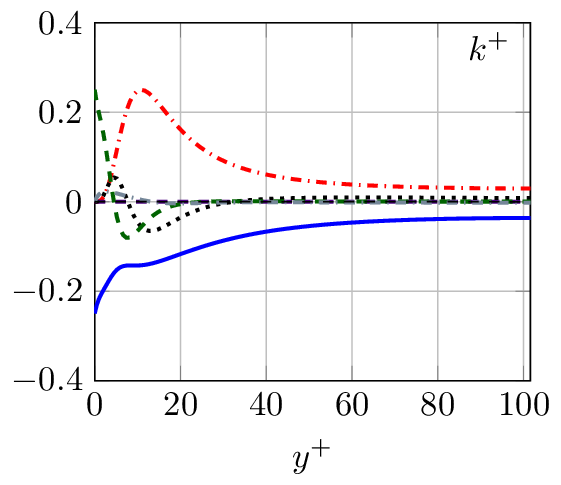}
\caption{Wall-normal profiles of the terms in the budget equation for $\aver{uu}^+$, $\aver{vv}^+$, $\aver{ww}^+$, $\aver{-uv}^+$ and $k^+$. The plots feature the turbulent transport (${\color{black}{\dotsb}}$), pressure transport (${\color[rgb]{0.46, 0.53, 0.6 }{\hdashrule[0.5ex]{1cm}{0.5mm}{2mm 3pt 0.5mm 2pt}}}$), viscous diffusion (${\color[rgb]{0, 0.39, 0}{\hdashrule[0.5ex]{1cm}{0.5mm}{2mm 3pt}}}$), production (${\color{red}{\hdashrule[0.5ex]{1cm}{0.5mm}{2mm 3pt 0.5mm 2pt}}}$), pressure strain (${\color[rgb]{0.29, 0, 0.50}{\hdashrule[0.5ex]{1cm}{0.5mm}{2mm 3pt}}}$) and pseudo-dissipation (${\color{blue}{\rule[0.5ex]{1cm}{0.5mm}}}$). The black dashed line plots the imbalance of each equation.}
\label{fig:singlepoint}
\end{figure}

Figure \ref{fig:profiles} shows the wall-normal profiles of the streamwise mean velocity and of the diagonal terms of the Reynolds stress tensor; they are compared with those computed by \cite{lee-moser-2018}. Considering that the Reynolds number is not identical, and that several other discretization choices differ (for example they employed a longer domain of $L_x = 20\pi h$ instead of $L_x = 12 \pi h$), the agreement is more than satisfactory. Figure \ref{fig:singlepoint} plots all the terms of the single-point budget for each non-zero component of the Reynolds tensor, and for the turbulent kinetic energy $k$. In each panel, the residual of the budget is also plotted: the imbalance due to the finite averaging time is negligible, with a maximum of less than $1.8\cdot 10^{-5}$.

The AGKE terms are computed from the database via a post-processing code that is derived with modifications from its Poiseuille counterpart described by \cite{gatti-etal-2020}. It also heavily relies on the numerical optimizations introduced by \cite{gatti-etal-2019} for a precursory GKE version, which computes correlations pseudo-spectrally whenever possible. For maximum accuracy, the derivatives in the homogeneous directions are computed in the Fourier space, whereas the derivatives in the wall-normal direction are evaluated by means of a finite-differences scheme with a five-points computational stencil. Since Couette and Poiseuille flows enjoy the same geometrical setting and in particular possess two homogeneous directions, the required modifications to the code from the Poiseuille version only involve changes to statistical symmetries. These changes are listed and discussed in Appendix \ref{sec:sym}. For validation, the residual of the AGKE balance equations is computed to ensure that statistical convergence is reached; it is indeed found to be negligible everywhere in the four-dimensional AGKE space.

\section{Results}
\label{sec:results}

Perhaps the most prominent feature of the turbulent Couette flow is the presence of elongated streamwise counter-rotating vortices which fill the gap between the moving walls. Before describing how these vortices affect the transport processes, the AGKE is used first to characterise these vortices. The analysis serves the dual purpose of discussing similarities and differences between Couette and Poiseuille flows, while demonstrating the effectiveness of the AGKE in discriminating quantitative characteristics of the turbulent structures. Figures comparing Poiseuille and Couette are presented when useful; the comparison is carried out with a turbulent Poiseuille flow at $Re_\tau=200$. Further details of this flow in the context of the AGKE can be found in \cite{gatti-etal-2020}. Note that such comparison is not unambiguous: owing to the different role of $Re$ in the two flows \citep[see for example][]{orlandi-bernardini-pirozzoli-2015, andreolli-quadrio-gatti-2021}, which Reynolds number should be kept fixed is not established, and the comparison must be intended in a qualitative sense only.

The AGKE equations are always computed in the four-dimensional space $(r_x,r_y,r_z,Y)$, but in this work results are mostly presented in the three-dimensional space with $r_x=0$. In fact, the streamwise separation $r_x$ is the least significant, because the large vortices are elongated in the $x$ direction, thus yielding the maximum correlation at $r_x=0$. An important mention of the $r_x \ne 0$ case is made in the concluding discussion in \S\ref{sec:conclusions}.

\subsection{The large-scale rolls described via the AGKE}
\label{sec:results-rolls}

\begin{figure}
\centering
\includegraphics[angle=-90,width=0.80\textwidth]{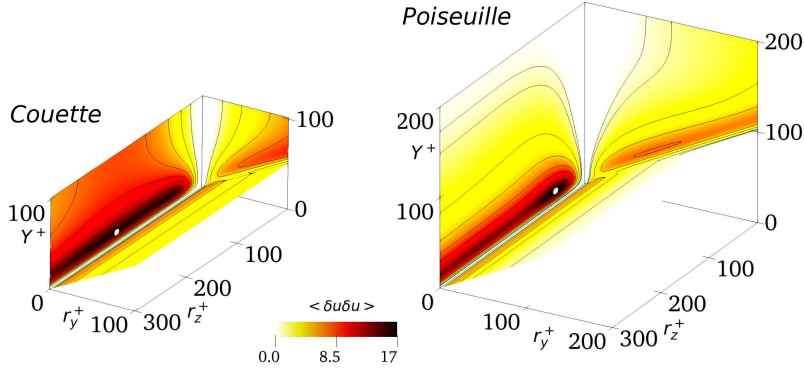}
\caption{Colormap of the structure function $\aver{\delta u \delta u}$ in the $r_x^+=0$ space, for Couette (left) and Poiseuille (right). The white marker locates the maximum of the structure function.}
\label{fig:ener}
\end{figure}

The quantitative characteristics of the large domain-filling streamwise rolls can be most easily identified by the structure function $\aver{\delta u \delta u}$, that is expected to present a maximum (i.e. largest negative correlation) at a spanwise scale $r_z$ corresponding to the spanwise vortex size. Figure \ref{fig:ener} shows a colormap for $\aver{\delta u \delta u}$ drawn on the bounding planes $r_y^+=0$, $r_z^+=0$ and $Y^+=r_y^+/2$ of a box in the $(r_z^+,r_y^+,Y^+)$ space, and compares Couette (left) and Poiseuille (right). The domain is bounded by the inclined plane $Y^+=r_y^+/2$ owing to  the separation $r_y$ being constrained by the presence of the wall, i.e. $r_y < 2 Y$. The presence of the Couette rolls with their large-scale energy footprint at the wall \citep{hwang-cossu-2010, lee-moser-2018} leads to a change of the position of the structure function maximum, which migrates towards larger spanwise separations. The position of the maximum increases from $r_z^+=59$ in Poiseuille to $r_z^+=183$ in Couette. The maximum region is for both flows located close to the wall at $Y^+ \approx 20$, and extends for a wide range of spanwise scales. In Couette, however, large values of $\aver{\delta u \delta u}$ are observed up to the channel centre; this is a striking difference with Poiseuille, and shows that the footprint of the rolls at the typical large spanwise scale $r_z^+ \approx 200$ or $r_z \approx 2h$ extends throughout the whole channel. 
Analogously, the statistical trace of the rolls can be clearly identified also in $\aver{\delta v \delta v}$, $\aver{\delta w \delta w}$ and $\aver{-\delta u \delta v}$ (not shown); Couette shows values of the structure functions significantly larger than $V_{ij}$ up to the channel core. Table \ref{tab:maxima} reports value and position of the maxima of each component of the structure function tensor for the Couette flow, together with those of other AGKE terms discussed below.

\begin{table}
\centering
\begin{tabular}{l c c c c c c c c} 
& \multicolumn{2}{c}{$\aver{\delta u_i \delta u_j}^+$} & \multicolumn{2}{c}{$\xi_{ij}^+$} & \multicolumn{2}{c}{$|\Pi_{ij}^+|$} & \multicolumn{2}{c}{$P_{ij}^+$} \\
          & value & position      & value & position    & value   & position  & value  & position  \\
$i=j=1$   & 16.3  & (0,183,15)    & 0.74  & (0,40,11)   & 0.20    & (0,50,20)  & 1.34   & (0,40,11) \\ 
$i=j=2$   & 2.9   & (0,100,101)    & 0.05  & (23,0,31)   & 0.11    & (0,40,3)  &  0     & -          \\
$i=j=3$   & 4.5   & (101,0,101)    & 0.09  & (0,40,7)   & 0.19    & (0,43,10)  &  0     & -          \\
$i=1,j=2$ & 2.5   & (0,133,101)    & 0.12  & (0,20,12)   & 0.25    & (0,66,15)  & 0.26   & (0,33,17)  \\
\end{tabular}
\caption{Maxima of $\aver{\delta u_i \delta u_j}^+$, source $\xi_{ij}^+$, absolute value of the pressure strain $|\Pi_{ij}^+|$ and production $P_{ij}^+$, and their positions in the $(r_y^+, r_z^+, Y^+)$ space for the Couette flow.}
\label{tab:maxima}
\end{table}

\begin{figure}
\centering
\includegraphics[angle=-90,width=0.80\textwidth]{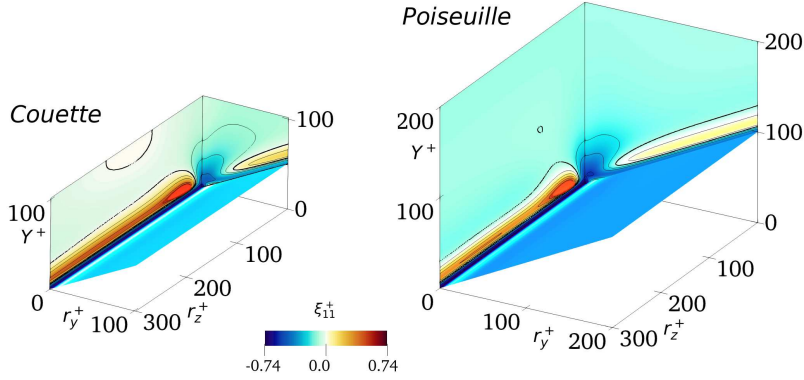}
\caption{Colormap of the source term $\xi_{11}^+$ for Couette (left) and Poiseuille (right). The colormap on the bounding planes, with the thick line indicating the contour level $\xi_{11}^+=0$, is plotted together with the isosurface $\xi_{11}^+=0.55$, which is $75 \%$ of the maximum value.}
\label{fig:source}
\end{figure}

\begin{figure}
\centering
\includegraphics[angle=-90,width=0.80\textwidth]{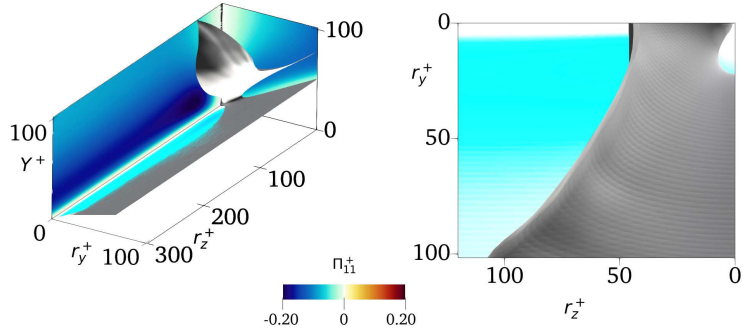}
\caption{Colormap of the pressure-strain term $\Pi_{11}^+$ for the Couette flow. The grey isosurfaces mark the boundary of non-preferential energy transfer from streamwise fluctuations towards the others. Inner side: $\Pi_{22}^+ > \Pi_{33}^+$. Outer side: $\Pi_{22}^+ < \Pi_{33}^+$. The right panel is a top view and zoom of the left panel.}
\label{fig:pstrain}
\end{figure}

It is interesting to ascertain whether scales and positions where the structure function is largest show net production $\xi_{11}>0$. Figure \ref{fig:source} plots coloured contours of the source term $\xi_{11}$ on the bounding planes, and an isosurface drawn in the volume at $75\%$ of the maximum. Of particular interest here is the contour line $\xi_{11}=0$, discriminating regions and scales where there is net gain of $\aver{\delta u \delta u}$ from those where there is net loss. Both flows present a region with net gain near the wall in correspondence of the buffer layer, i.e. $Y^+ \approx 14$. $\xi_{11}$ peaks in this region at $(r_y^+,r_z^+) \approx (0,40)$ indicating that in both cases the largest production of turbulent fluctuations is associated with the near-wall regeneration cycle \citep{jimenez-pinelli-1999}. However, Couette features a second region with net positive production (of lower intensity) at the channel centre, at a spanwise scale of $r_z^+ \approx 150$. Since the pressure strain term $\Pi_{11}$, shown in figure \ref{fig:pstrain}, and the dissipation $D_{11}$ (not shown) are negative (almost) everywhere, the local maximum of $\xi_{11}$ at the centreline is unequivocally related to the positive production of the large rolls. Indeed, this positive source in the channel core, albeit of limited intensity because of the relatively low $Re$, implies that production $P_{11}^+$ is larger than the sum of pressure strain $\Pi_{11}^+$ and dissipation $D_{11}^+$ for $r_z^+ \sim 150$. This production at the channel centre is absent in Poiseuille, where the anti-symmetry of the mean velocity gradient rules out any production. Interestingly, in Poiseuille a second tiny positive region can be identified, which is further from the wall but not at the centreline. Although based on this plot alone it might be interpreted as statistical noise, it is known \citep{gatti-etal-2020} that this is an early appearance of the outer turbulence cycle, which becomes more and more evident as $Re$ increases, but does not extend up to the centreline.

A key contributor to the source term is the redistribution provided by the pressure-strain term: figure \ref{fig:pstrain} plots the $\Pi_{11}^+$ component as an example. It is (slightly) positive at all scales near the wall, but it is negative everywhere else: the streamwise fluctuations loose energy to feed the cross-stream components at all scales for $Y^+>2$. The AGKE visualise at which scales and positions $\aver{\delta u \delta u}$ is preferentially redistributed towards either $\aver{\delta v \delta v}$ or $\aver{\delta w \delta w}$. In figure \ref{fig:pstrain} a grey isosurface is drawn to mark the boundary between the two situations. Following \cite{gatti-etal-2020}, the isosurface is defined by $\Pi_{22}^+ /\Pi_{11}^+ = \Pi_{33}^+ /\Pi_{11}^+ = - 0.5$, with $\Pi_{11}^+ < 0$. In the fraction of the volume delimited by the gray isosurface towards the $Y^+$ axis, it is found that $\Pi_{22}^+ /\Pi_{11}^+ < - 0.5$, hence $\Pi_{11}^+$ preferentially redistributes streamwise fluctuations towards the wall-normal fluctuations. This occurs up to the channel centre, where it is true for all $r_y^+$ and $r_z^+ < 100$ (see the zoomed right panel in figure \ref{fig:pstrain}). On the remaining volume, instead, the preferential receiver of streamwise fluctuations energy is the spanwise component $\aver{\delta w \delta w}$. A similar picture was already observed in Poiseuille flow \citep{gatti-etal-2020}, although here the preferential redistribution to $\aver{\delta v \delta v}$ takes place within a narrower range of separations.

Note that the structure functions and the net productions peak at similar but not identical positions (see table \ref{tab:maxima}). These differences will be discussed in the following.

\subsection{The diagonal components of $\aver{\delta u_i \delta u_j}$}
\label{sec:results-diagonal}

We now proceed to study the interscale transfer processes in turbulent Couette flow. As a general observation, the transfers described by \cite{gatti-etal-2020} for the Poiseuille flow are visible here too: they look nearly identical for $\aver{\delta u \delta u}$ and extremely similar for the other diagonal components. They attest the universality of the near-wall-turbulence cycle for the small scales. Hence, the following discussion will be mostly focused on bringing out the role of the large streamwise rolls, which are absent in Poiseuille.

\begin{figure}
\centering
\includegraphics[angle=-90,width=0.45\textwidth]{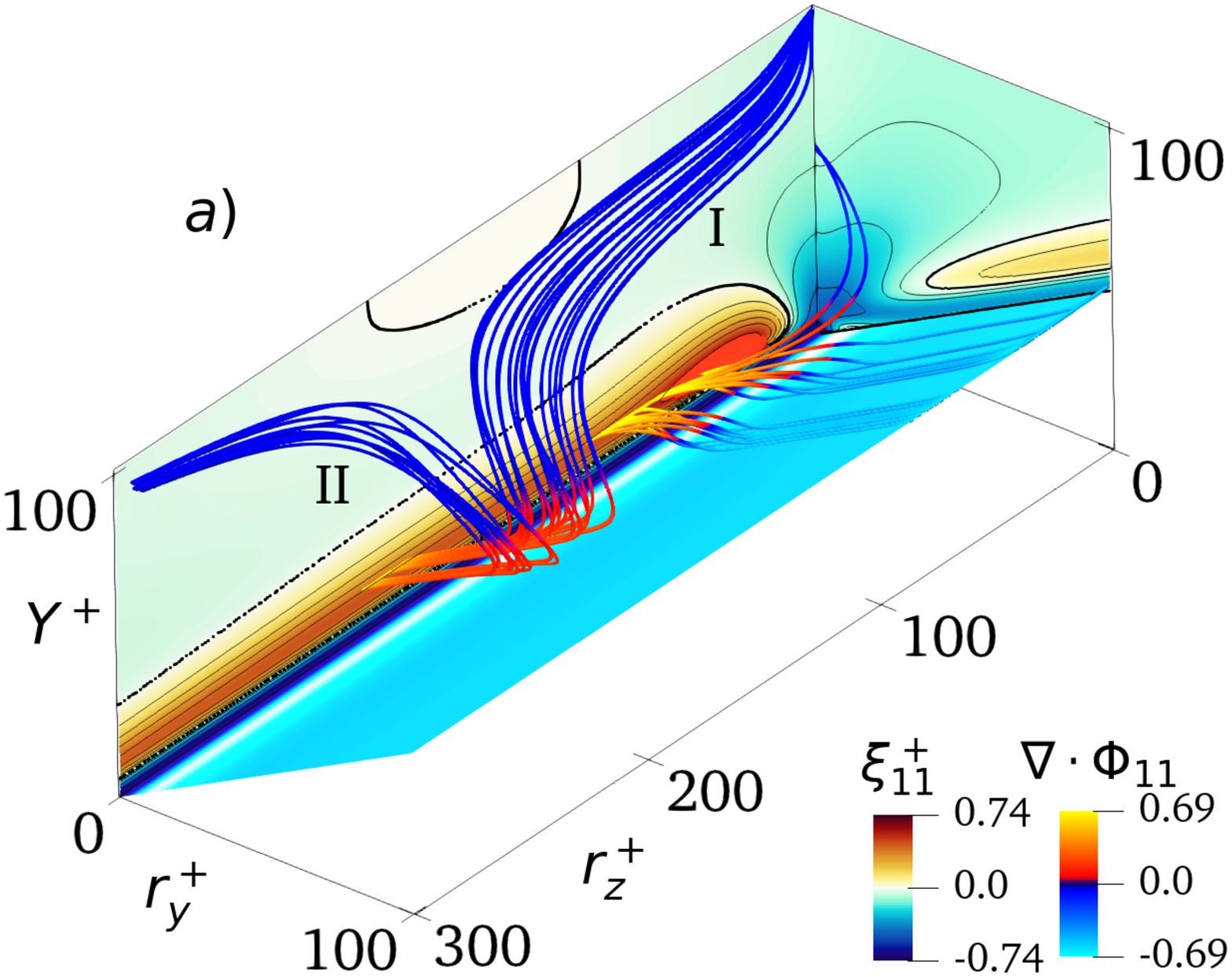}
\includegraphics[angle=-90,width=0.45\textwidth]{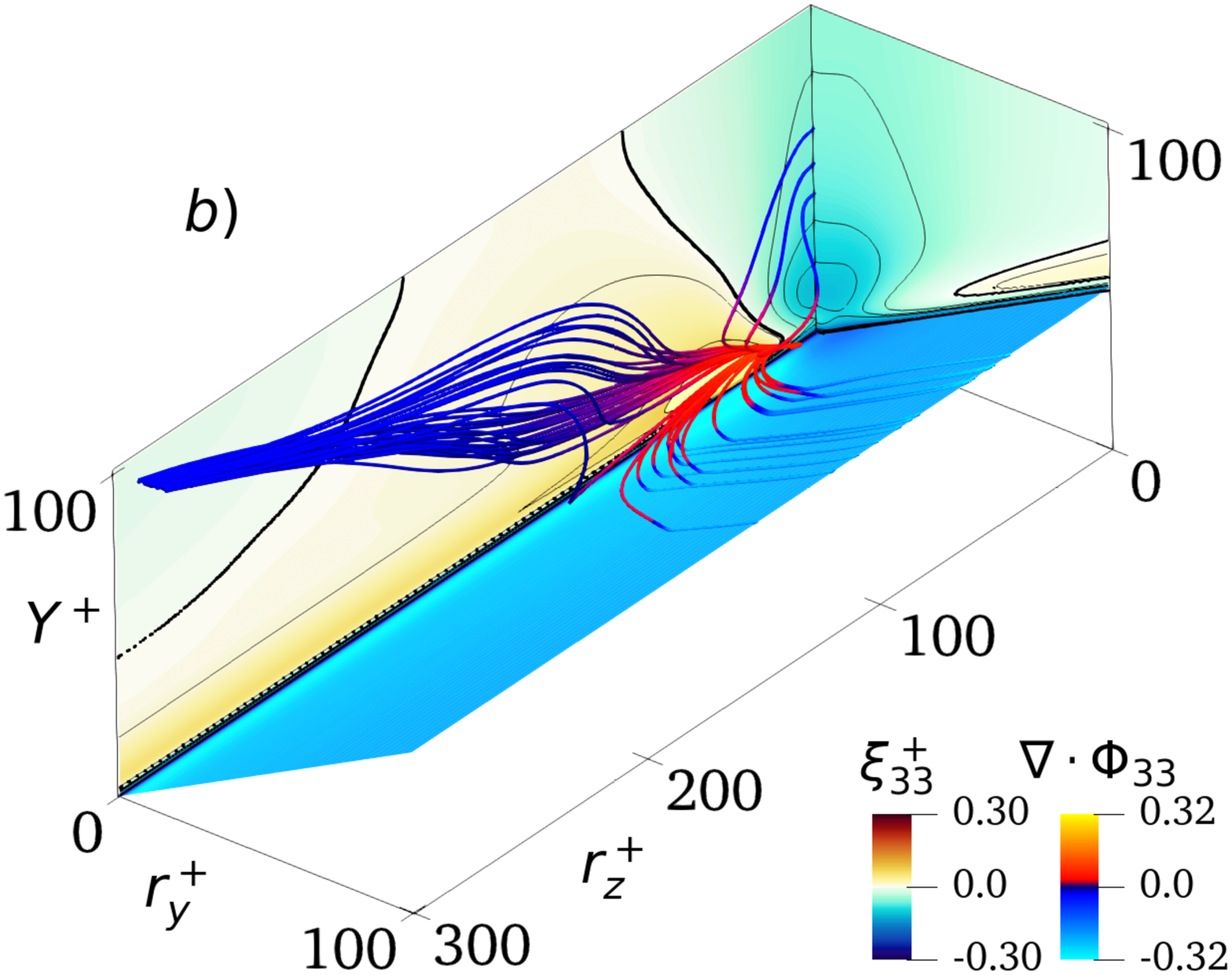}
\includegraphics[angle=-90,width=0.45\textwidth]{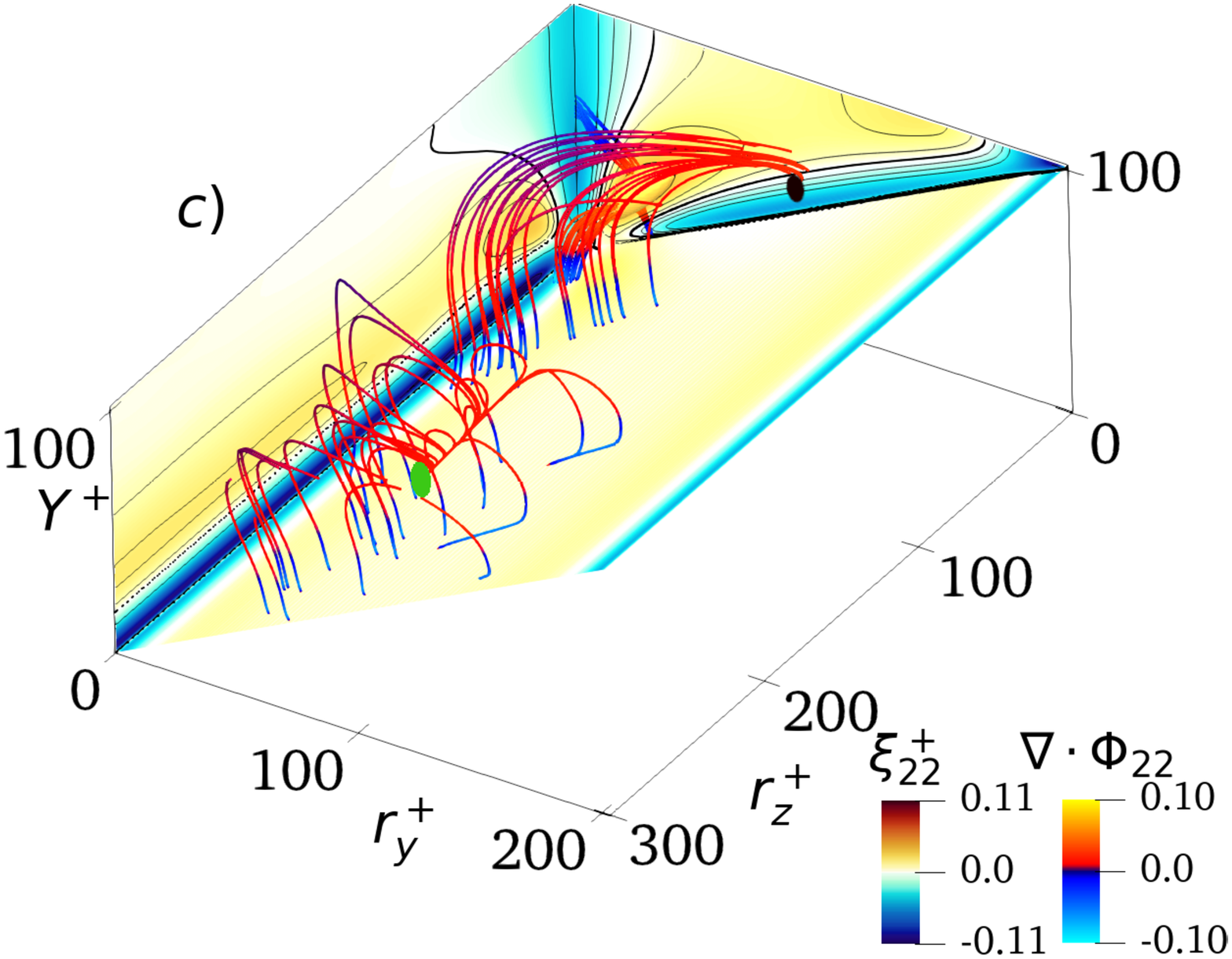}
\caption{Field lines of the vector of the fluxes $\vect{\Phi}=(\phi_y,\phi_z,\psi)$ coloured by its divergence $\vect{\nabla \cdot \Phi}=\partial \phi_y/\partial r_y + \partial \phi_z/\partial r_z + \partial \psi / \partial Y$. Panel (a): $\aver{\delta u \delta u}$; panel (b): $\aver{\delta w \delta w}$; panel (c): $\aver{\delta v \delta v}$. Colour contours of the corresponding source term are shown on the bounding planes, with a black thick line indicating the zero level. The isosurfaces correspond to $75\%$ of the source maximum, i.e. $\xi_{11}^+=0.55$, $\xi_{22}^+= 0.035$, and $\xi_{33}^+=0.07$. In panel (c), the black dot is the origin of the field lines; the green dot is the singularity point from which other field lines originate.}
\label{fig:flux}
\end{figure}

\begin{figure}
\centering
\includegraphics[width=0.8\textwidth]{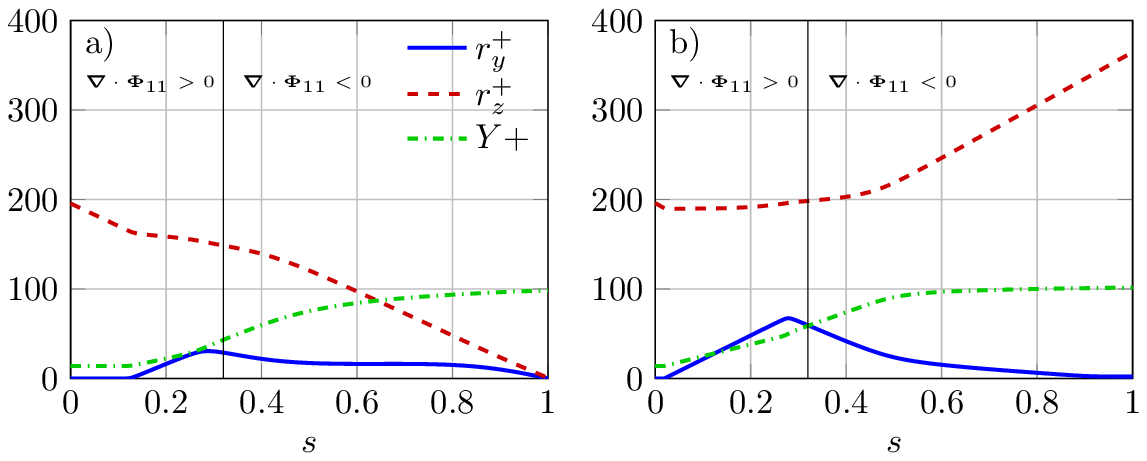}
\includegraphics[width=0.8\textwidth]{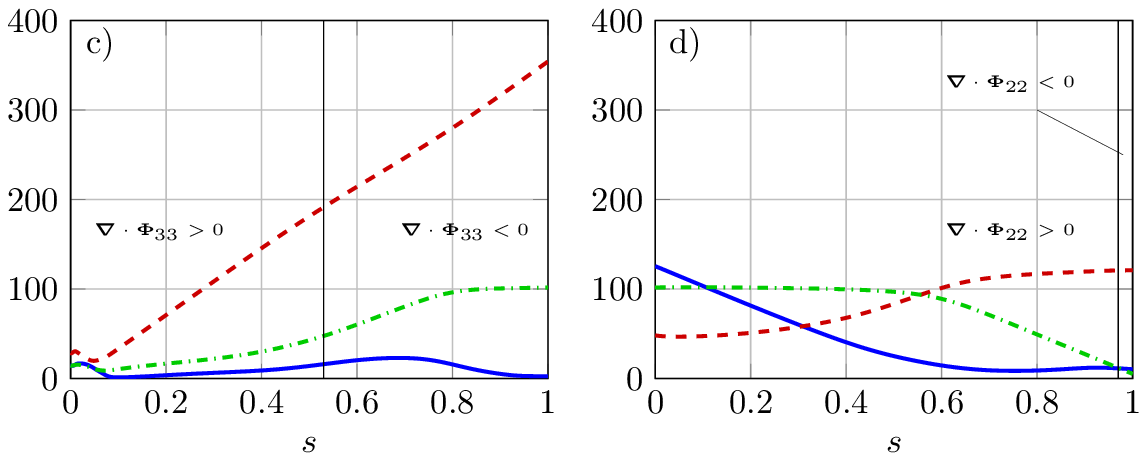}
\caption{Evolution of the values of $r_y^+ ( {\color{blue}{\rule[0.5ex]{1cm}{0.5mm}}})$, $r_z^+ ( {\color{red!80!black}{\hdashrule[0.5ex]{1cm}{0.5mm}{2mm 3pt}}})$ and $Y^+ ( {\color{green!80!black}{\hdashrule[0.5ex]{1cm}{0.5mm}{2mm 3pt 1mm 2pt}}}) $ along a representative field line for $\aver{\delta u \delta u}$ set I (a) and set II (b); (c) for $\aver{\delta w \delta w}$ and (d) for $\aver{\delta v \delta v}$. The dimensionless arch length $s$ is defined as $s= 1/s_{max} \int_{0}^{s_{max}}ds$, with $ds = \sqrt{dr_y^2 + dr_z^2 + dY^2}$. The black vertical lines mark the position where the divergence changes sign along the line.}
\label{fig:field_lines}
\end{figure}


The availability of fluxes in the AGKE enables a precise description of the transfer processes through their field lines, that visualise how the fluctuations are transferred among scales and positions. Although energy is not necessarily bound to be continuously transported along these lines, and the flow dynamics may be instantaneously very complex, the fluxes explain the different scales and positions at which $\aver{\delta u_i \delta u_j}$ and $\xi_{ij}$ have their peak, and their field lines visualise and help understanding their spatial arrangement. Figure \ref{fig:flux} plots the field lines of the flux vector $\vect{\Phi}=(\phi_y,\phi_z,\psi)$ for $\aver{\delta u \delta u}$, $\aver{\delta v \delta v}$ and $\aver{\delta w \delta w}$. The lines are coloured by their divergence, i.e. $\vect{\nabla \cdot \Phi} = \partial \phi_y/\partial r_y + \partial \phi_z/\partial r_z + \partial \psi / \partial Y$, and the colourmap shown in the background is for the corresponding source. While the field lines convey directional information, their colour code is meant to provide quantitative information about their energetic relevance: a positive value of $\vect{\nabla \cdot \Phi}$ indicates fluxes energised by local mechanisms, while a negative value indicates fluxes releasing energy. Indeed, large positive values of $\vect{\nabla \cdot \Phi}$ are observed in the vicinity of the points where the lines originate, whereas large negative values in the regions where they vanish. 

The fluxes of $\aver{\delta u \delta u}$ are considered first; see panel (a) of figure \ref{fig:flux}. The major contributors to these fluxes are (not shown) the viscous and turbulent transport, with comparable intensity. The field lines of the total flux originating close to the maximum of $\xi_{11}$ are analogous to those of the Poiseuille case, and show the same type of transfer; hence they are not discussed here \citep[details can be found in][]{gatti-etal-2020}. However, other field lines point to an additional transfer mechanism at work in Couette, which exchanges $\aver{\delta u \delta u}$ between the near-wall structures and the large streamwise rolls. These lines originate near the wall but just beneath the source maximum in the core, i.e. at $(r_y^+,r_z^+,Y^+) \approx (0,200,14)$, at a location in agreement with the peak of the spectral transfer of $\aver{uu}$, placed by \cite{kawata-alfredsson-2019} at $r_z^+=160$ and $Y^+=18$ in a Couette flow at $Re_\tau=63$. The lines show that part of the excess of $\aver{\delta u \delta u}$ produced in the near-wall region is transferred towards the channel centre to feed scales and positions associated with the large rolls, indicating a  bottom-up interaction. This is highlighted by $\vect{\nabla \cdot \Phi}_{11}$ which indicates that the fluxes are energised for $Y^+ \le 30$ but release energy at larger distances. The rolls are therefore fed by both their own production and the wall turbulence itself. The field lines, once the channel core is reached, are repulsed by the region of positive source. Some lines (collectively referred to as set I) deviate towards the smallest spanwise scales. Others (set II) reach larger $r_z$. Panels (a) and (b) of figure \ref{fig:field_lines} describe quantitatively how  $r_y$, $r_z$ and $Y$ evolve along a field line of each set, in terms of the normalised position $s$ along the line. In the same figure the position(s) along the lines where the divergence of the flux is zero is shown, thus identifying regions where energy is released or received. The excess of $\aver{\delta u \delta u}$ associated with the spanwise rolls is dissipated at both the smallest scales via the usual direct cascade (set I) and at the large uncorrelated motions via a mixed cascade (set II), in which $r_y$ decreases while $r_z$ increases. The sink where the lines of set I end is mainly due to viscous effects, whereas along lines of set II energy is lost owing to a combined effect of pressure-strain and viscosity.

The mutual connection between the near-wall region and the channel core is also evident by looking at the other wall-parallel component $\aver{\delta w \delta w}$. Like $\xi_{11}$, the source $\xi_{33}$ peaks in the near-wall region at $(r_y^+,r_z^+,Y^+)=(0,40,7)$ and has large positive values up to the channel centre at $r_z^+ \approx 100 $ (see figure \ref{fig:flux}b). However, in this case the positive source is due to the redistribution of the streamwise component of the turbulent energy via the pressure-strain term, since the production $P_{33}$ term is zero as $\partial W / \partial y=0$. Unlike for $\aver{\delta u \delta u}$, all the field lines of $\aver{\delta w \delta w}$ originate close to the maximum of the source, i.e. at $(r_y^+,r_z^+,Y^+)=(12,26,13)$. They are mainly determined (not shown) by the turbulent contribution, the viscous flux being smaller in magnitude. Again, some lines resemble those of Poiseuille vanishing at the smallest scales along the $Y$ axis, but others provide a bottom-up transfer of $\aver{\delta w \delta w}$ from the smaller near-wall structures to the large spanwise rolls. Indeed, panel (c) of figure \ref{fig:field_lines} shows that part of the excess of $\aver{\delta w \delta w}$ near the wall first feeds the large spanwise rolls via an inverse cascade (see the first part of the lines where both $r_z$ and $r_y$ increase) and then dissipates at $r_y \rightarrow 0$ and large $r_z$ via a mixed cascade ($r_y$ decreases and $r_z$ increases); note again that $\vect{\nabla \cdot \Phi}_{33}$ is positive in the near-wall region and negative away from the wall where the fluxes release energy to the large-scale structures. Overall, a bottom-up transfer of $\aver{\delta w \delta w}$ is observed together with an inverse cascade followed by a mixed cascade.

Finally, $\xi_{22}$ too peaks in the near-wall region for scales and positions associated with the near-wall cycle, i.e. $(r_y^+,r_z^+,Y^+)=(24,0,31)$, and it is positive up to the channel centre with a local maximum at $(r_y^+,r_z^+,Y^+) \approx (120,0,100)$ (see figure \ref{fig:flux}c). As for $\aver{\delta w \delta w}$, positive values of $\xi_{22}$ mean that $\Pi_{22}$ is positive and larger than $|D_{22}|$ since the production term of $\aver{\delta v \delta v}$ is null. Therefore, the pressure strain feeds the vertical fluctuations at both the small near-wall scales and the large scales away from the wall, yielding a net production of $\aver{\delta v \delta v}$. Field lines at the small scales near the wall are analogous to the Poiseuille case. Lines intercepting larger scales, instead, clearly show a top-down transfer, with scales away from the wall feeding the near-wall structures. The pressure spatial transfer dominates the flux, although viscous scale transfer and turbulent scale transfer play a role at small and large scales respectively. Some lines originate at $(r_y^+,r_z^+,Y^+)=(130,60,101)$, a point marked by the black dot in figure \ref{fig:flux}c. These fluxes are further energised by the excess of $\aver{\delta v \delta v}$ in the region with $r_y \rightarrow 0$ and larger $r_z$, as indicated by the the positive $\vect{\nabla \cdot \Phi}_{22}$. Then they release $\aver{\delta v \delta v}$ in the near-wall region ($\vect{\nabla \cdot \Phi}_{22}<0$) and eventually vanish at the wall where $\xi_{22}<0$ due to the negative contribution of both the dissipation and (mainly) the pressure strain. The negative $\Pi_{22}$ in the vicinity of the wall, together with the positive $\Pi_{11}$ (see figure \ref{fig:pstrain}) and $\Pi_{33}$ (not shown), indicates that the vertical fluctuations turn into wall-parallel ones owing to the splatting effect \citep{mansour-kim-moin-1988} present in all wall-bounded flows. Other lines originate from a singularity point at $(r_y^+,r_z^+,Y^+)=(120,280,100)$ (green dot in the same figure) where again $\xi_{22}>0$. These lines are straight at first with $Y$ and $r_y$ remaining constant (although figure \ref{fig:flux}c only shows those going towards lower $r_z$), with further energised fluxes; then they reorient towards the wall and become nearly vertical, i.e. $r_y$ and $r_z$ remain constant, as they are attracted by the sink at the wall. Before vanishing at the wall these fluxes release $\aver{\delta v \delta v}$ in the near-wall region over a wide range of scales. Overall, the excess of $\aver{\delta v \delta v}$ produced away from the wall at large $r_y$ due to the pressure-strain action feeds a wide range of scales in the near-wall region, where $\aver{\delta v \delta v}$ is dissipated due to the combined action of pressure-strain and viscosity.

\begin{figure}
\centering
\includegraphics[angle=-90,width=0.8\textwidth]{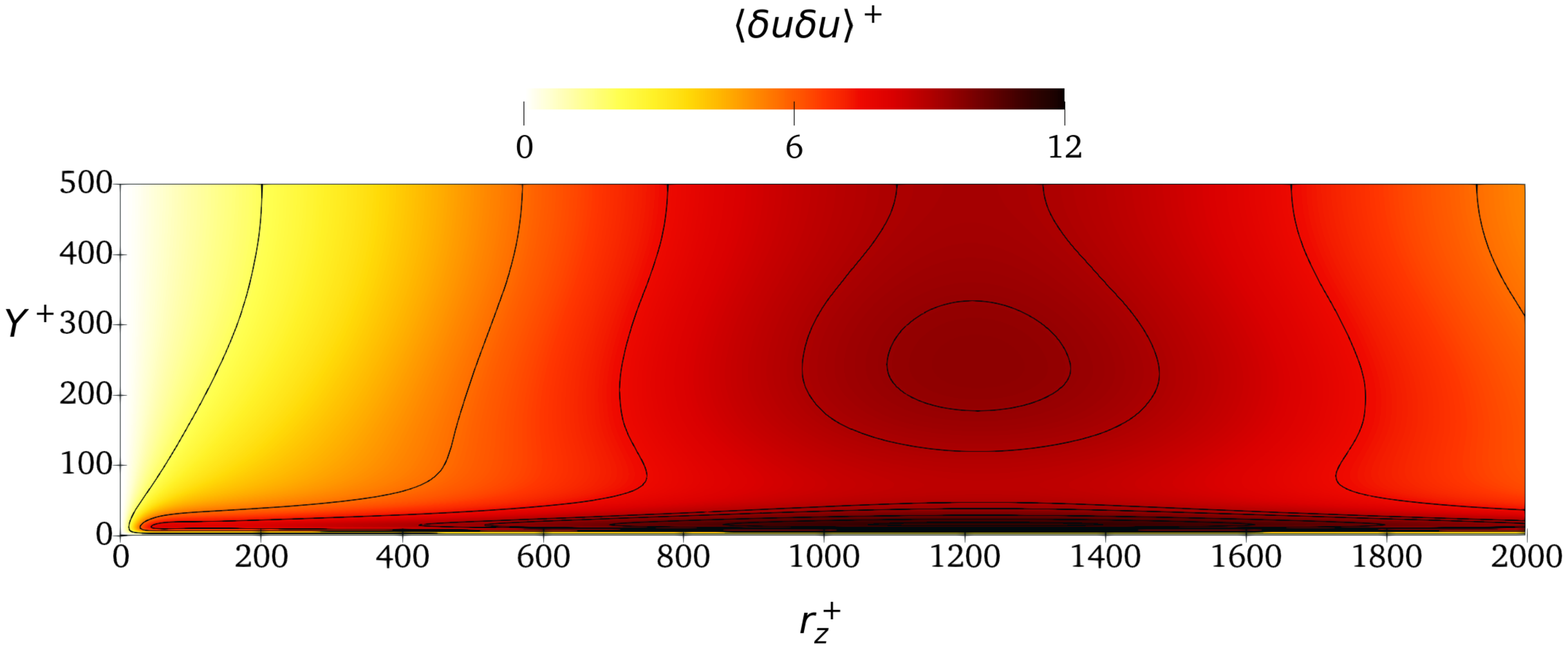}
\includegraphics[angle=-90,width=0.8\textwidth]{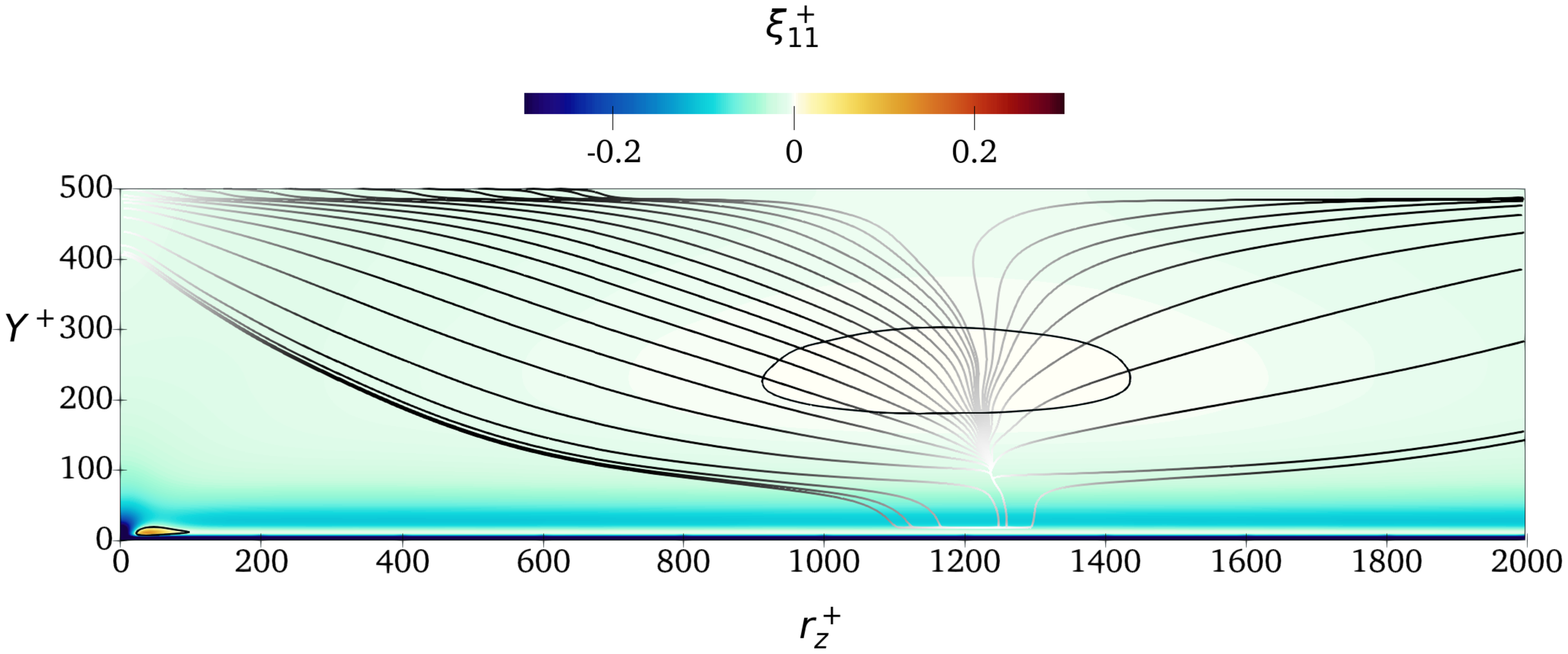}
\caption{Top: $\aver{\delta u \delta u}$ at $Re_\tau=500$ at $r_x=r_y=0$. Bottom: field lines of the fluxes of $\aver{\delta u \delta u}$ coloured with their magnitude, with colormap of the source term shown in the background (black thick lines indicate the zero level).}
\label{fig:retau500}
\end{figure}

Although studying how these statistics change with increasing values of $Re$ is not the primary concern of this paper, it can be anticipated that the $Re$ effects do not qualitatively alter the general picture, since $Re_\tau=100$ is already large enough for  the outer structures to develop. To confirm this, we have processed a dataset at $Re_\tau=500$ taken from \cite{andreolli-quadrio-gatti-2021}; as an example, results for the $\aver{\delta u \delta u}$ component are provided in figure \ref{fig:retau500}. The top panel plots the structure function itself. Compared to fig.\ref{fig:ener} (left), the near-wall peak is now accompanied by a visible outer peak at $r_z \approx 2h$, and the contours still extend to the centerline. The lower panel, to be compared to the plane at $r_y=0$ in fig.\ref{fig:flux}a, portraits fluxes and the source term. The field lines still originate near the wall at $r_z \approx 2h$, and proceed towards larger $Y$ to give birth to set I and II. The near-wall maximum of the source is connected to the near-wall cycle and thus scales in wall units, whereas the outer maximum becomes more evident. The source term at low $Re$ is dominated by production and its outer maximum is exactly at the centerline, whereas at $Re_\tau=500$ the source maximum associated to the outer cycle is found at $Y^+ \approx 250$: the large-scale production keeps peaking at the centerline (not shown), but the source term also contains non-negligible contributions from pressure strain, and has its maximum around the scale and the wall-normal distance of the outer cycle.  

In summary, the AGKE reveals the existence of an inverse energy cascade for $\aver{\delta u \delta u}$ and $\aver{\delta w \delta w}$, coupled with an ascending spatial transport that spans the whole domain. On the contrary, $\aver{\delta v \delta v}$, which is governed by the redistribution term $\Pi_{22}$, shows a top-down transport. As discussed later in \S\ref{sec:conclusions}, this picture partially differs from the description by \cite{kawata-alfredsson-2018}, where the diagonal components of the Reynolds stress tensor reportedly present a top-down transfer with a direct cascade only.  

\subsection{The off-diagonal component $\aver{-\delta u \delta v}$}
\label{sec:results-offdiagonal}

\begin{figure}
\centering
\includegraphics[angle=-90,width=0.90\textwidth]{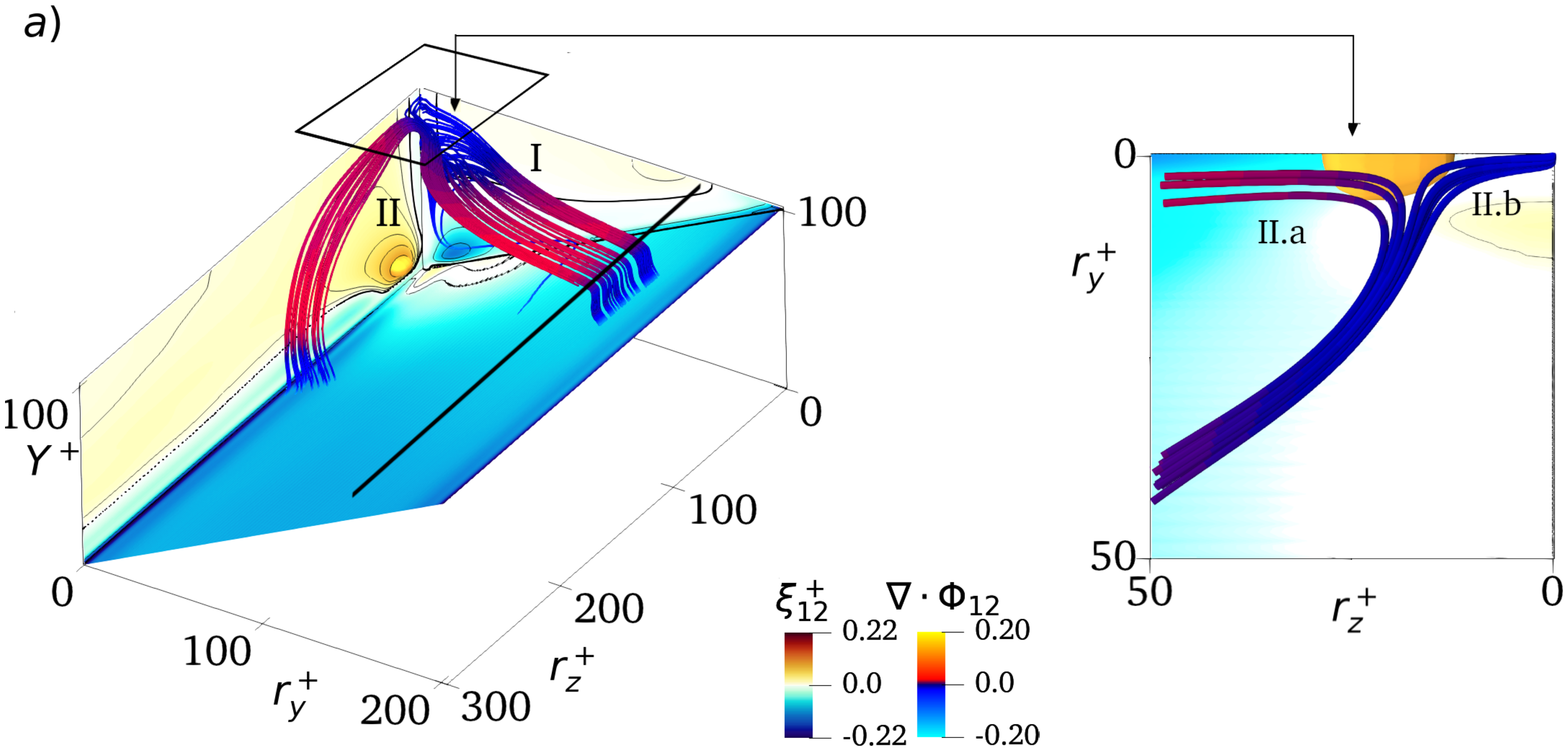}
\includegraphics[width=0.95\textwidth]{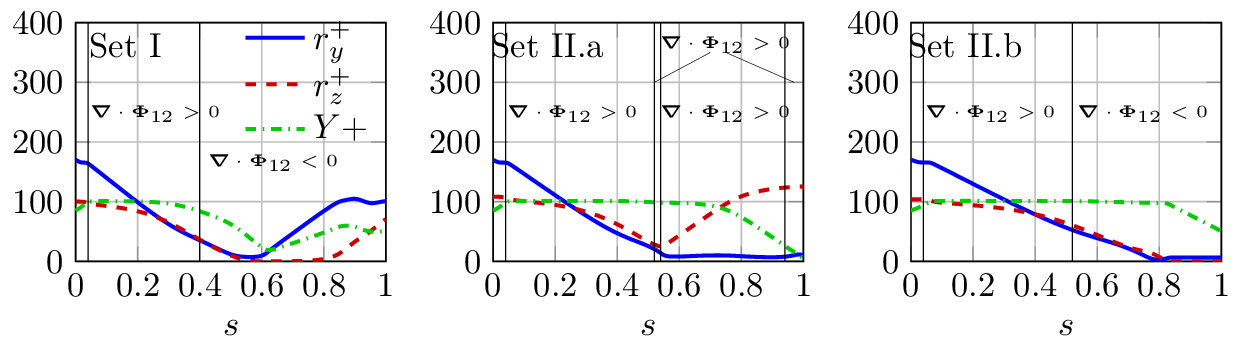}
\caption{Off-diagonal component $\aver{-\delta u \delta v}$. Top: field lines of the vector of the fluxes $(\phi_{y,12}, \phi_{z,12}, \psi_{12})$, coloured by the their divergence and grouped in sets I and II. Colour contours of the source term $\xi_{12}^+$ are shown on the bounding planes, with a black thick line indicating the zero contour level. The black straight line placed at $Y^+ = 85$ and $r_y^+=170$ indicates all the spanwise separations from which some field lines originate. The yellow isosurface corresponds to $\xi_{12}^+= 0.09$ The top right panel is a zoomed top view of the lines of set II. The evolution of $r_y^+ ( {\color{blue}{\rule[0.5ex]{1cm}{0.5mm}}})$, $r_z^+ ( {\color{red!80!black}{\hdashrule[0.5ex]{1cm}{0.5mm}{2mm 3pt}}})$ and $Y^+ ( {\color{green!80!black}{\hdashrule[0.5ex]{1cm}{0.5mm}{2mm 3pt 1mm 2pt}}}) $ along a representative field line from sets I, II.a and II.b are shown in the lower panels, in which the black vertical lines mark the position where the divergence changes sign along the line.}
\label{fig:dudv_flux}
\end{figure}
  
The interaction between the large-scale dynamics and the near-wall turbulence is also observed in the off-diagonal component of the structure function tensor $\aver{-\delta u \delta v}$. Like the single-point quantity $\aver{-uv}$, this component is not defined in its sign, hence $\aver{-\delta u \delta v}$ should not be interpreted in terms of energy. The transfer of $\aver{- \delta u \delta v}$ is shown in figure \ref{fig:dudv_flux}a, where field lines of the fluxes are plotted together with the source term $\xi_{12}$. Fluxes taking place at scales and positions associated with the near-wall cycle are similar to Poiseuille and are not discussed here; for clarity the associated lines are not shown in figure \ref{fig:dudv_flux}a. The remaining field lines are mainly determined (not shown) by the pressure spatial transport, although scale turbulent transport is comparable. Turbulent transport and viscous diffusion are the only non-zero contributions in the scale space, the latter dominating at the small scales and the former at the large scales. The flux lines originate from a straight line placed at large wall-normal distances ($Y^+ = 85$ and $r_y^+=170$), where $\xi_{12}$ is positive, and involve all spanwise separations (to simplify figure \ref{fig:dudv_flux}a, field lines are only  plotted for a range of spanwise separations, and a black straight line indicates where they originate). These lines are bound to be tangent to the plane $Y^+=h^+$, because the symmetries of the terms involved in the budget equation for $\aver{- \delta u \delta v}$ imply $\psi_{12} = 0$ at the centreline for $r_x^+ = 0$ (see Appendix~\ref{sec:sym}). Starting from this straight line, some field lines bend and descend vertically towards the wall attracted by the sink $\xi_{12}<0$. On the contrary, other field lines direct first towards smaller vertical scales and then descend towards the wall. Overall, $\aver{-\delta u \delta v}$ is transferred from the central region, where it is maximum, towards the wall region, indicating again that the large rolls interact with the near-wall structures. 

The lines going towards smaller $r_y$ can be further divided into two topologically distinct groups, named set I and set II. While reducing their $r_y$, the two line sets diverge from each other. Those of set I go towards smaller $r_z$ and smaller $Y$ being attracted by the region with negative source at $r_z^+=0$ (see the dark blue region in figure \ref{fig:dudv_flux}a); at $r_z^+ \rightarrow 0$ and small $r_y$ they collapse onto a single line that first sticks to the plane $r_z^+=0$ and then turns to become a well defined line at $Y^+=50$ and $r_z^+=100$. While descending towards the wall, they release $\aver{-\delta u \delta v}$ as indicated by the negative values of $-\vect{\nabla \cdot \Phi}_{12}$. Conversely, while reducing their $r_y$, lines of set II are repulsed by the positive peak of $\xi_{12}^+$ on the $r_y^+=0$ plane. Some lines remain at large $Y^+$ and release $\aver{-\delta u \delta v}$ at the smallest scales, ending up on the vertical axis corresponding to zero separations. Most of them, however, release $\aver{-\delta u \delta v}$ at larger $r_z^+$ and lower $Y^+$ and eventually vanish at the wall.

The lower panels in fig.\ref{fig:dudv_flux} help understanding how separations and the wall-normal position change along selected lines of each set. For set I, the transport of $\aver{- \delta u \delta v}$ at first follows the classical energy cascade: the spanwise separation decreases while the transfer takes place from the channel centre towards the wall, until $r_z^+ \approx 0$ is reached. In a later stage, $r_y^+$ and $Y^+$ increase again, and eventually after the turn they become locked at $r_y^+=100$ and $Y^+=50$, while $r_z^+$ increases. For set II, for which a zoom is available in fig.\ref{fig:dudv_flux}b, the lines of subset II.b simply direct towards the $Y^+$ axis to dissipate. Those of subset II.a, instead, show first a continuous decrease of both $r_y^+$ and $r_z^+$, then the wall-normal separation remains constant at $r_y^+ \approx 7$ as $r_z^+$ increases, while $Y^+$ initially remains constant and reduces afterwards to reach zero value. On such lines an inverse cascade is observed, albeit in a limited range of scales, namely $5 < r_y^+ < 10$ and $20 < r_z^+ < 135$, while a spatial transfer starts from the channel centre and proceeds towards the wall.

\section{Concluding discussion}
\label{sec:conclusions}

\begin{table}
\centering
\begin{tabular}{c|cccc} 
                           & bottom-up  & top-down & direct &  inverse \\
$k$                        & yes  &     & yes   &  yes \\
$\aver{\delta u \delta u}$ & yes  &     & set I &  set II \\
$\aver{\delta v \delta v}$ &      & yes &       &  yes  \\
$\aver{\delta w \delta w}$ & yes  &     &       &  yes  \\
$\aver{\delta u \delta v}$ &      & yes & set I &  set II.a \\
\end{tabular}
\caption{Types of transport observed in a turbulent Couette flow.}
\label{tab:sintesi}
\end{table}

The transfer of Reynolds stresses has been considered component-wise in a turbulent Couette flow by means of the Anisotropic Generalized Kolmogorov Equations or AGKE, which provide a complete and quantitative description of direct and inverse cascades in the space of scales as well as transfers in the wall-normal direction, for each component of the Reynolds stress tensor. The field lines of the fluxes of Reynolds stresses naturally visualize cascading in physical and scale spaces. When fluxes are observed along the wall-normal direction, top-down and bottom-up processes can be discerned; conversely, when fluxes across scales are considered, direct and inverse cascades are observed. The study builds upon a DNS database, produced for this study at $Re_\tau = 101.6$, which matches the largest $Re$ considered in the experimental study of \cite{kawata-alfredsson-2018}. What makes the Couette flow particularly interesting in the present context is its non-zero mean velocity gradient at the centreline, which enables turbulence production, and the presence of large-scale spanwise rolls, whose interaction with the smaller turbulent structures is not entirely understood.

The turbulent Couette flow shows a rich scenario, where direct and inverse cascades coexist, together with bottom-up and top-down interactions. While some of these patterns are also present in plane Poiseuille flow, Couette features some peculiarities, which have been addressed by the present study.

The statistical trace of the large rolls is clearly observed in the structure function $\aver{\delta u \delta u}$. In particular, $\aver{\delta u \delta u}$ is significantly larger than the local covariance in a region encompassing a well-defined spanwise scale of $r_z^+ \approx 200$ or $r_z \approx 2h$ throughout the whole channel height, with local maxima in the buffer layer and in the core. At the channel centre the source term $\xi_{11}$ is positive and quantifies the contribution of the large-scale structures to the overall production; by increasing $Re$ the maximum of the source moves towards the wall, because of the relative increased importance of the pressure-strain redistribution, which negatively contributes to the source. 

Studying the field lines of the flux vector $\vect{\Phi}=(\phi_y, \phi_z, \psi)$ brings to light the transfer processes among scales and positions. The two wall-parallel components, $\aver{\delta u \delta u}$ and $\aver{\delta w \delta w}$, are mainly transferred from the large and small scales near the wall towards larger scales away farther from it. The excess of energy produced near the wall 
partially feeds the large spanwise rolls, pointing to a scenario of bottom-up transport. In terms of cascade, $\aver{\delta u \delta u}$ features both direct and inverse cascades, whereas the fluxes of $\aver{\delta w \delta w}$ reveal the coexistence of an inverse and a mixed cascade. The $\aver{\delta v \delta v}$ component behaves differently, as expected, and shows a top-down transfer where both inverse and mixed cascades occur. The surplus of $\aver{\delta v \delta v}$ at the channel core, supplied by the pressure-strain term $\Pi_{22}$, is transferred towards the near-wall region via two mechanisms. In the first, small $r_z$ scales away from the wall feed near-wall eddies with larger spanwise but lower wall-normal scales. In the second, large $r_z$-scale motions placed at the channel core feed near-wall motions over a wide range of scales. The off-diagonal component $\aver{- \delta u \delta v}$  mainly features a top-down transfer with both direct and inverse cascades, although the inverse one is restricted to a confined range of $r_y^+$ and $r_z^+$. The dominant transport processes in both the space of scales and in the physical space are summarised in Table \ref{tab:sintesi}.

This picture of interscale transfers confirms the conclusions of \cite{kawata-alfredsson-2018}, who found an influence of the small near-wall scales upon the large scales away from the wall. However, the present AGKE analysis offers more insight, thanks to its ability to precisely discriminate the various processes and the scales and positions at which they take place. Some of the extra details brought forward here for the first time are not in agreement with the existing broad picture. 
\cite{kawata-alfredsson-2018} pointed out that the production of turbulent kinetic energy at large scales is supported by the small near-wall scales by means of the Reynolds shear stress that is transferred from small to large scales throughout the channel. Moreover, they observed a transfer of turbulent kinetic energy from the large scales away from the wall to the small near-wall scales. While the general trend of near-wall scales feeding the large structures at the channel core is confirmed, the present results reveal different space- and scale-transfers and therefore different sustaining mechanisms. We reveal that the Reynolds shear stress $\aver{-\delta u \delta v}$ is mainly transferred from the large scales in the channel core to the near-wall region, where it is released at both small and large scales. On the contrary, for the turbulent kinetic energy we have described a complex picture, with a bottom-up transfer for the wall-parallel components $\aver{\delta u \delta u}$ and $\aver{\delta w \delta w}$ and a top-down transfer for the vertical one $\aver{\delta v \delta v}$. Thus, the present scenario indicates that the large-scale structures away from the wall are maintained by their production mechanism and by the transfer of the $u$- and $w$- energy contributions from the large and small near-wall scales respectively. In turn, the large-scale motions away from the wall support the near-wall motions, as shown by the transfers of $\aver{\delta v \delta v}$, and their production mechanism through the top-down transfer of $\aver{-\delta u \delta v}$. 

\cite{kawata-tsukahara-2021}, based on their spectral analysis, provided a schematic model to explain the self-sustaining cycle of the large vortices in the Couette flow. The current AGKE analysis confirms the essence of their model, while bringing in some additional details. The model by \cite{kawata-tsukahara-2021} starts by observing that $\aver{uu}$ energy is present at the large scales: this is seen by the positive production at large streamwise scales (figure 10 in their paper), which descends from a positive shear coupled with the positive $\aver{-uv}$ produced by the large-scale $\aver{vv}$ (its origin is explained below). Then $\aver{uu}$ is transferred towards smaller scales as shown by the interscale-transfer term (figure 11 in their paper), following a mechanism described as destabilization of the streaks. At the small scales, the pressure-strain mechanism redistributes $\aver{uu}$ towards $\aver{vv}$ and $\aver{ww}$, and feeds the small-scale streamwise vortices (figure 12 in their paper). Finally, $\aver{ww}$ is transferred from the small scales to the large scales. Here the pressure strain redistributes $\aver{ww}$ towards $\aver{vv}$, leading to a production for $\aver{-uv}$ and, consequently, for $\aver{uu}$ at the large-scales. The loop is thus closed.

\begin{figure}
\centering
\includegraphics[width=0.9\textwidth]{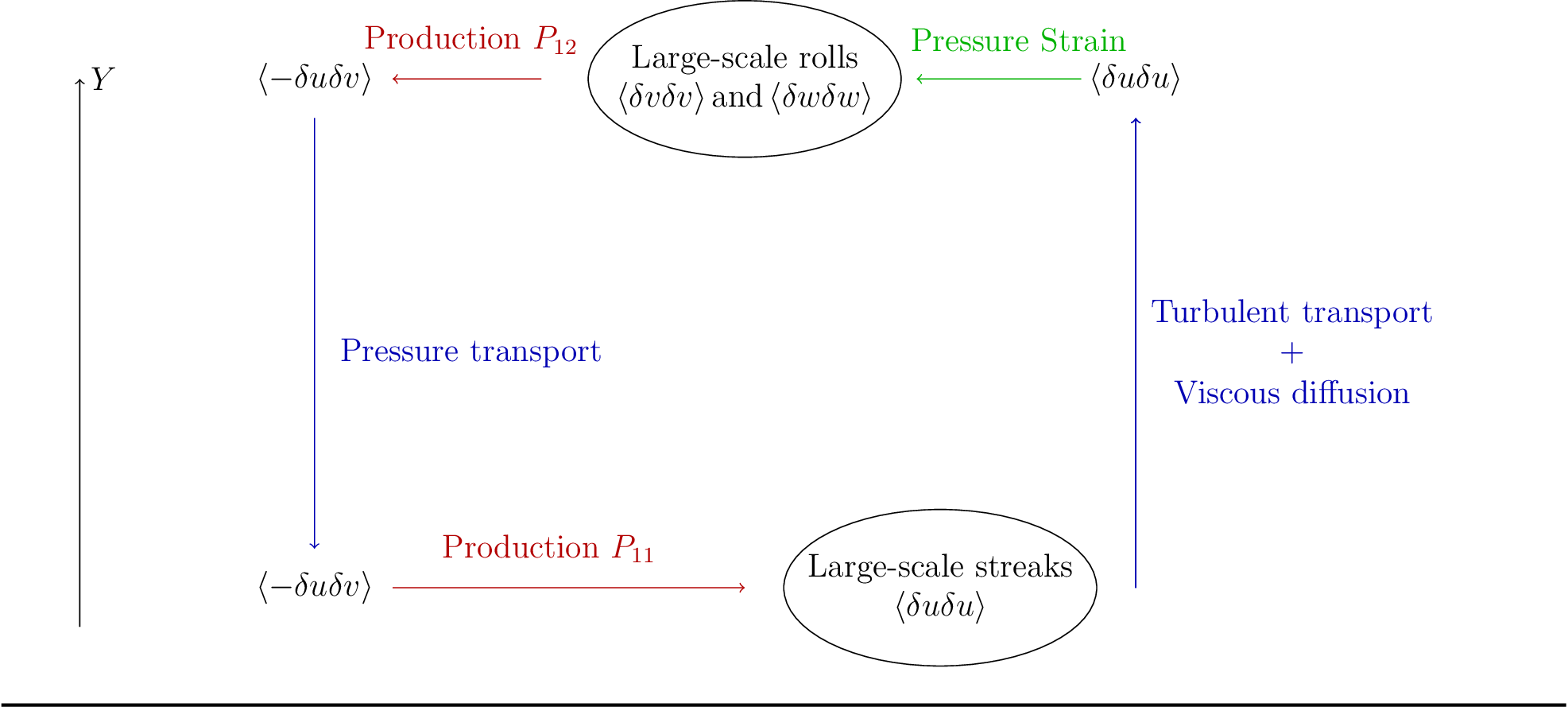}
\caption{Schematic model of the closed-loop sustaining mechanism for $\aver{\delta u_i \delta u_j}$ in the $r_x=0$ space, associated with the large-scale motions.}
\label{fig:model}
\end{figure}

Our analysis is consistent with this view, yet the picture is slightly different and enriched. Figure \ref{fig:model} schematically shows the model of the closed-loop self-sustaining mechanism based on the AGKE results. The main difference is that with the AGKE the transfer among scales at different $Y$ becomes observable, so that the model can be described as follows. Large $\aver{\delta u \delta u}$ energy found at $r_z^+ \approx 180$ and $Y^+ \approx 15$ is the statistical trace of large-scale streaks (see figure \ref{fig:ener}). They are fed by the mean shear and by the positive $\aver{-\delta u \delta v}$, associated with the large-scale rolls and released from the fluxes at these large spanwise separations close to the wall. The excess of $\aver{\delta u \delta u}$ associated with the streaks is transferred to both larger and smaller scales, as well as both towards and away from the wall, by the combined action of viscous diffusion and turbulent transport. The most part is transferred away from the wall (see figure \ref{fig:flux}a), where pressure strain redistributes $\aver{\delta u \delta u}$ towards the cross-stream components. Hence, these transfers and the pressure-strain activity explain how the large-scale streaks feed the large-scale vortices. The fluxes of $\aver{\delta w \delta w}$ (figure \ref{fig:flux}b) indicate that the spanwise fluctuations of the large spanwise rolls are fed not only by the pressure-strain redistribution, but also by the near-wall cycle. 
At large $Y$ and large $r_z$ the large $\aver{\delta v \delta v}$ together with the mean shear generate positive production for $\aver{-\delta u \delta v}$. The produced $\aver{-\delta u \delta v}$ is then transferred towards the wall via pressure transport, and part is released by the fluxes at $r_z ^+ \approx 150$ (family II.a in figure \ref{fig:dudv_flux}). The resulting positive $\aver{-\delta u \delta v}$, joint with the mean shear, is responsible for $P_{11}>0$ thus sustaining the large-scale streaks. The loop is thus closed.

\begin{figure}
\centering
\includegraphics[angle=-90,width=0.49\textwidth]{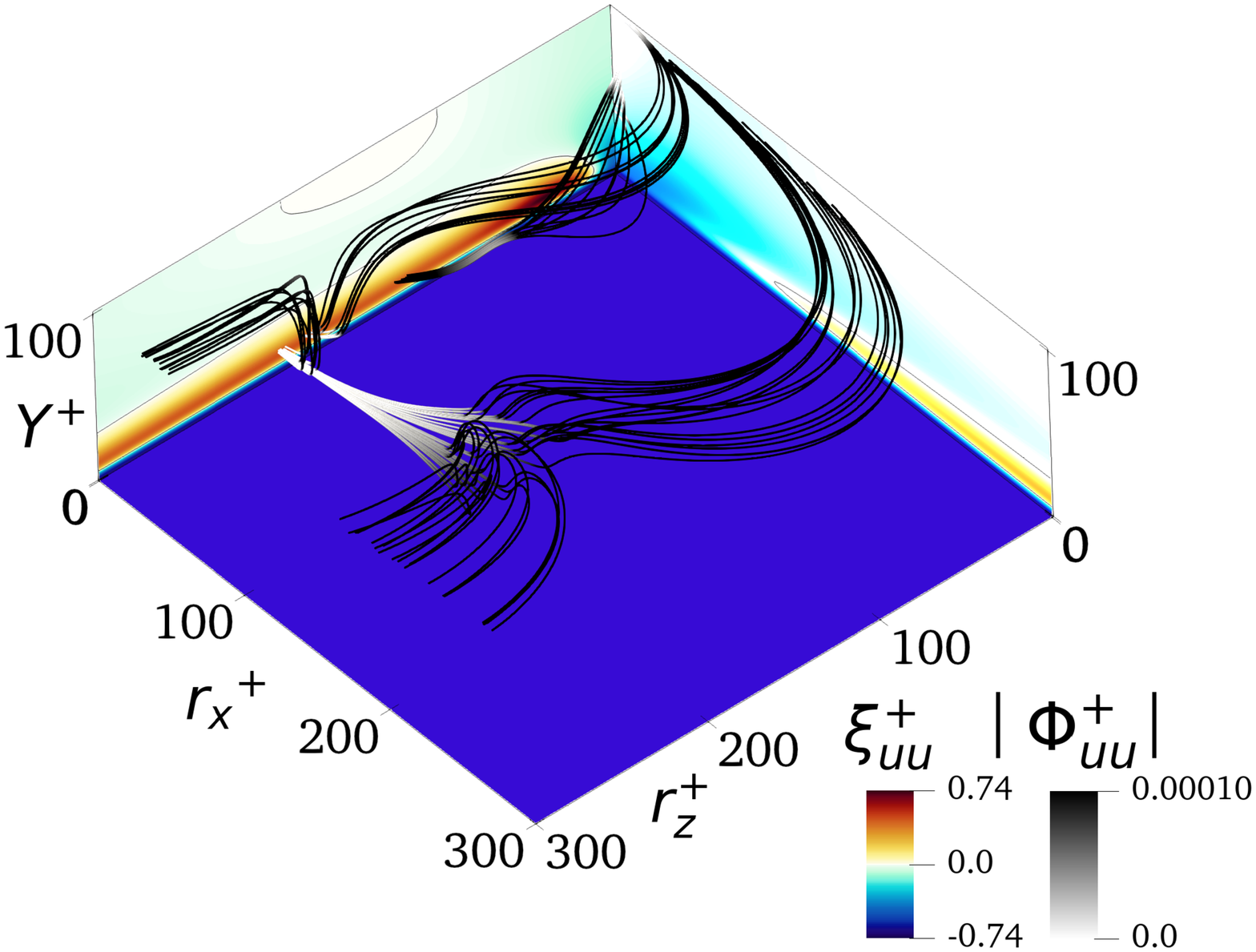}
\includegraphics[angle=-90,width=0.49\textwidth]{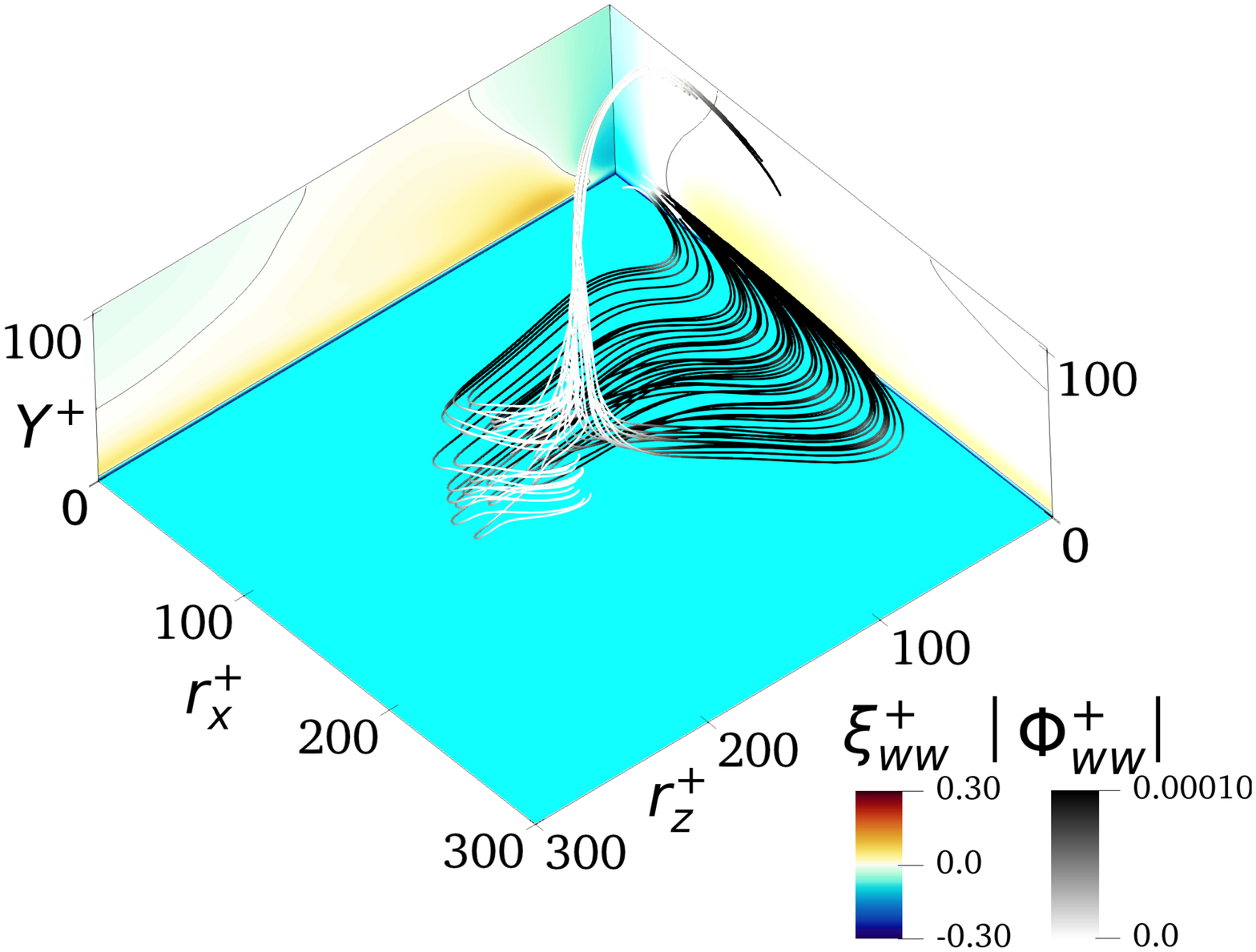}
\caption{Field lines of the vector of the fluxes in the $(r_x, r_z, Y)$ space, for $\aver{\delta u \delta u}$ (left) and $\aver{\delta w \delta w}$ (right). Colour contours of the corresponding source term are shown on the bounding planes. The flux lines are coloured with the flux magnitude.}
\label{fig:rx}
\end{figure}

Although so far the streamwise separation has not been considered, it must be said that concerns have been recently raised on the conclusions derived from such partial observations neglecting streamwise separations/wavenumbers. For example, \cite{kawata-tsukahara-2021} found via a spanwise-Fourier-mode analysis that an inverse transfer of $\aver{uu}$ is still present when the computational domain is restricted in its spanwise size to remove the outer structures. Hence, such mechanism would not represent an interaction between the near-wall cycle and the large outer structures, as confirmed by the absence of an inverse cascade in the analysis of the streamwise Fourer modes. They suggested that this may be related to the self-sustaining cycle of the inner and/or outer structures. To address this point, we show in figure \ref{fig:rx} fluxes for $\aver{\delta u \delta u}$ and $\aver{\delta w \delta w}$ in the three-dimensional space at $r_y=0$ that considers the streamwise separation too. 

For $\aver{\delta u \delta u}$ the lines associated to the large scales originate at $r_x=0$, thus their starting point is the same as in the $r_x=0$ space. They develop both close to the wall and at larger $Y$; streamwise energy is transferred via both direct and inverse cascades. These lines can thus be interpreted, as done in the previous analysis with $r_x=0$, as part of the self-sustaining cycle of the large-scale rolls. The lines of $\aver{\delta w \delta w}$ confirm that they initially possess small spanwise separations (much like in the channel flow) and then move towards large $r_z$ and large $Y$, indicating again an interaction of the small scales of the near-wall cycle with the large spanwise rolls located away from the wall. Thus, looking at $r_x = 0$ already seems to provide the correct picture, and suggests that $\aver{\delta u \delta u}$ is driven by an autonomous large-scale mechanism and the transfers of $\aver{\delta w \delta w}$ describe an interaction between the near-wall cycle and the large outer structures.

The present analysis has mostly considered a single, relatively low value of the Reynolds number, and thus its extension to higher $Re$ remains interesting and useful. Nevertheless, already at this $Re$, thanks to their capability to properly define the concept of scale in the wall-normal direction and to provide a detailed view of each component of the Reynolds stress tensor, the AGKE have been essential to understand and describe the interaction between the near-wall turbulence and the elongated rolls of the turbulent Couette flow.

\section*{Acknowledgments}
The authors acknowledge computing support by the state of Baden-Württemberg through bwHPC. D.G. acknowledges partial support by the Priority Programme SPP 1881 Turbulent Superstructures of the Deutsche Forschungsgemeinschaft.

\section*{Funding} 
This research received no specific grant from any funding agency, commercial or not-for-profit sectors.

\section*{Declaration of Interests} 
The authors report no conflict of interest.

\bibliographystyle{jfm}

\appendix
\section{Symmetries}
\label{sec:sym}

The symmetries in the terms of the AGKE equations are listed below for the case of the plane Couette flow. 

The terms appearing in the budget equations of $\aver{\delta u \delta u}$, $\aver{\delta v \delta v}$, $\aver{\delta w \delta w}$ and $\aver{\delta u \delta v}$ are first considered. The inversion of $\vect{r}$ and of the spanwise coordinate $z$ lead to the same symmetries as in the Poiseuille flow \citep{gatti-etal-2020}. $\vect{r} \rightarrow -\vect{r}$ leads to $\vect{\phi} \rightarrow -\vect{\phi}$, $\psi \rightarrow \psi$, $\xi \rightarrow \xi$ and $\aver{\delta u_i \delta u_i} \rightarrow \aver{\delta u_i \delta u_i}$; $z \rightarrow -z$ implies $r_z \rightarrow -r_z$ and consequently $\phi_x \rightarrow \phi_x$, $\phi_y \rightarrow \phi_y$, $\phi_z \rightarrow - \phi_z$, $\psi \rightarrow \psi$, $\xi \rightarrow \xi$ and $\aver{\delta u_i \delta u_i} \rightarrow \aver{\delta u_i \delta u_i}$. On the contrary, the inversion of the wall-normal coordinate $y$ leads to different symmetries compared to the Poiseuille flow, as it implies also the inversion of the streamwise coordinate $x$. It leads to $Y \rightarrow -Y$, $r_x \rightarrow -r_x$, $r_y \rightarrow -r_y$ and consequently $\phi_x \rightarrow - \phi_x$, $\phi_y \rightarrow - \phi_y$, $\phi_z \rightarrow \phi_z$, $\psi \rightarrow - \psi$, $\xi \rightarrow \xi$ and $\aver{\delta u_i \delta u_i} \rightarrow \aver{\delta u_i \delta u_i}$.

The terms involved in the budget equations of $\aver{\delta u \delta w}$ and $\aver{\delta v \delta w}$ are now considered. In comparison to the other components, their symmetries are the same for an inversion of $\vect{r}$ but differ when the inversions of $z$ or $y$ (and $x$) are considered. When $z$ is inverted the AGKE terms undergo $\phi_x \rightarrow - \phi_x$, $\phi_y \rightarrow - \phi_y$, $\phi_z \rightarrow \phi_z$, $\psi \rightarrow - \psi$, $\xi \rightarrow - \xi$ and $\aver{\delta u_i \delta u_i} \rightarrow  -\aver{\delta u_i \delta u_i}$, as in the Poiseuille flow. The inversion of $y$ (and $x$) leads to $\phi_x \rightarrow \phi_x$, $\phi_y \rightarrow \phi_y$, $\phi_z \rightarrow - \phi_z$, $\psi \rightarrow \psi$, $\xi \rightarrow - \xi$ and $\aver{\delta u_i \delta u_i} \rightarrow  -\aver{\delta u_i \delta u_i}$ for $\aver{\delta u \delta w}$ and $\aver{\delta v \delta w}$.

Overall, these symmetries imply that
some terms of the AGKE are zero in particular regions of the four-dimensional domain. These regions (except the trivial case of $\vect{r}=\vect{0}$) are listed below for each component $\aver{\delta u_i \delta u_j}$. For notational simplicity the origin of the wall-normal coordinate is shifted to the centreline of the channel.
 
For $\aver{\delta u \delta u}$, $\aver{\delta v \delta v}$, $\aver{\delta w \delta w}$ and $\aver{\delta u \delta v}$:

\begin{center}
  \begin{align*}
    \phi_x(Y,0,0,r_z) &= 0 \\
    \phi_y(Y,0,0,r_z) &=0 \\
    \phi_z(Y,r_x,r_y,0) & = 0 \\
    \psi(0,r_x,r_y,r_z) &=0.    
  \end{align*}
\end{center}

For $\aver{\delta u \delta w}$ and $\aver{\delta v \delta w}$:

\begin{center}
  \begin{align*}
    \phi_x(Y,r_x,r_y,0) &=0 \\
    \phi_y(Y,r_x,r_y,0) &=0 \\
    \phi_z(Y,0,0,r_z) &=0 \\
    \psi(Y,r_x,r_y,0) &=0 \hspace{2cm} & \psi(Y,0,0,r_z)=0& \\
    \psi(0,r_x,r_y,r_z) &=0 \\    
    \xi(Y,r_x,r_y,0) &=0               & \xi(Y,0,0,r_z)=0& \\
    \aver{\delta u_i \delta u_j}(Y,r_x,r_y,0) &=0               & \aver{\delta u_i \delta u_j}(Y,0,0,r_z)=0&.
  \end{align*}
\end{center}

\end{document}